\documentclass[aps,prd,12pt]{revtex4}
\usepackage{epsfig}
\usepackage{graphicx}
\usepackage[font=small,skip=0pt,justification=raggedright]{caption}
\usepackage{amsmath}
\usepackage{amssymb}
\usepackage{mathrsfs}
\usepackage{verbatim}
\usepackage{dutchcal}

\usepackage[normalem]{ulem}
\usepackage{xcolor}

\newcounter{fig}   \newcommand{\lbfig}[1]{\refstepcounter{fig}
\label{#1} }

\newcommand{\bea}{\begin{eqnarray}}
\newcommand{\eea}{\end{eqnarray}}
\newcommand{\be}{\begin{equation}}
\newcommand{\ee}{\end{equation}}

\newcommand{\re}[1]{(\ref{#1})}

\newcommand{\cgg}{\mathcal{g}}

\newcommand{\eqn}{\begin{eqnarray}}
\newcommand{\eqnx}{\end{eqnarray}}

\date{\today}

\begin{document}
\title{$U(1)$ gauged boson stars in the\\ Einstein-Friedberg-Lee-Sirlin model}

\author{J.~Kunz}
\affiliation{Institute of Physics, University of Oldenburg,
Oldenburg D-26111, Germany}
\author{V.~Loiko}
\affiliation{ Department of Theoretical Physics and Astrophysics,
Belarusian State University, Minsk 220004, Belarus}
\author{Ya.~Shnir}
\affiliation{BLTP, JINR, Dubna 141980, Moscow Region, Russia
%Department of Theoretical Physics, Tomsk State %Pedagogical University, Russia
%Institute of Physics,
%Carl von Ossietzky University Oldenburg, Germany
%Oldenburg D-26111, Germany
}

\begin{abstract}
We consider spherically symmetric $U(1)$ gauged boson stars in the two-component scalar Friedberg-Lee-Sirlin model with a symmetry breaking potential in 3+1 dimensional spacetime.
Depending on the relative strength of gravity and the electromagnetic interaction, the resulting boson stars exhibit either the typical properties of ungauged boson stars, or their behavior resembles the pattern found for gauged Q-balls of the  Friedberg-Lee-Sirlin model in flat spacetime, both for a finite and a vanishing potential.
\end{abstract}
\maketitle

%%%%%%%%%%%%%%%%%%%%%%%%%%%%%%%%%%%%%%%%%%%%%%%%%%%%%%%%%%
\section{Introduction}
%%%%%%%%%%%%%%%%%%%%%%%%%%%%%%%%%%%%%%%%%%%%%%%%%%%%%%%%%%

Q-balls represent time-dependent non-topological solitons with a stationary oscillating phase
\cite{Rosen,Friedberg:1976me,Coleman:1985ki}.
They may exist in models with a complex scalar field in Minkowski spacetime possessing an unbroken, continuous global symmetry (for reviews, see, e.g. \cite{Lee:1991ax,Radu:2008pp,Shnir2018}).
Q-balls carry a Noether charge associated with this symmetry.
This charge is proportional to the angular frequency of the complex scalar field and can be interpreted as the particle number of the Q-balls. 

Q-balls arise in a variety of models. Important
examples are theories with a sextic potential \cite{Friedberg:1986tq,Volkov:2002aj,Kleihaus:2005me,Kleihaus:2007vk}, 
supersymmetric extensions of the Standard Model \cite{Kusenko:1997zq},
and coupled two-component systems of a complex scalar field and a real scalar field with symmetry breaking potential (Friedberg-Lee-Sirlin (FLS) model) \cite{Friedberg:1976me}.
Further, there are gauged Q-balls in models with local $U(1)$ symmetry
\cite{Lee:1988ag,Lee:1991bn,Kusenko:1997vi,Anagnostopoulos:2001dh,Gulamov:2015fya,Gulamov:2013cra,Panin:2016ooo,Nugaev:2019vru}.
The presence of the electromagnetic interaction affects the properties of the gauged Q-balls. 
In particular, they may exist only for a restricted range of values of the gauge coupling.

When gravity is coupled to the stationary oscillating scalar field, the field configurations can be stabilized even in the simplest case of a massive Einstein-Klein-Gordon theory, and the corresponding solutions are now commonly referred to as boson stars
\cite{Kaup:1968zz,Feinblum:1968nwc,Ruffini:1969qy,Colpi:1986ye,Deppert:1979au}. 
In models with self-interacting potentials there are then solitonic boson stars 
\cite{Friedberg:1986tq,Kleihaus:2005me,Kleihaus:2007vk}, that are smoothly linked to the corresponding Q-balls in Minkowski spacetime. 
Properties of charged boson stars, including investigation of their stability and critical behavior
were studied in \cite{Jetzer:1989us,Jetzer:1992tog,Pugliese:2013gsa,Kleihaus:2009kr,Kumar:2014kna}.

Here we consider charged boson stars in the $U(1)$-gauged two-component model in (3+1)-dimensional asymptotically flat spacetime, i.e., the Einstein-Maxwell-Friedberg-Lee-Sirlin (EMFLS) model. 
The model thus describes a self-gravitating coupled system of a complex scalar field, minimally coupled to the Maxwell field, and a real scalar field. This theory may serve as a toy-model to study field configurations localized by gravity in more realistic theories with with symmetry breaking potential, like the Standard Model.
We investigate the influence of the presence of the U(1) charge on the boson stars, obtained previously within the Einstein-Friedberg-Lee-Sirlin (EFLS) model in
\cite{Kunz:2019sgn}. We show that  distinctive new features of the gauged boson stars in the EMFLS model
are related to the delicate force balance between gravitational attraction, electrostatic repulsion and the short and long range scalar interactions.

The paper is organized as follows: We introduce the model in section 2, where besides the action and the equations of motion we also discuss the Ansatz for the metric and the fields and the boundary conditions for the functions. 
In section 3 we present the results obtained by solving the coupled system of equations numerically. 
We first address the limit without gravity, i.e., the Q-ball limit,
and demonstrate the influence of charge. 
We then couple gravity and study the dependence on the strength of the gravitational coupling constant as well as the dependence on the strength of electromagnetic coupling constant.
Moreover we address the limit when the mass of the real scalar field vanishes such that it becomes long-ranged.
We close with our conclusions in section 4.

\section{The model}

We now present the Einstein-Maxwell-Friedberg-Lee-Sirlin model, describing a self-gravitating coupled system of a complex scalar field $\phi$, minimally interacting with an Abelian gauge field $A_\mu$, and a real scalar field $\psi$. 
The corresponding action is given by
\be S=\int d^4 x \sqrt{-\cgg} \left(\frac{R}{4\alpha^2} - L_{m}\right),
\label{lag}
\ee
where the rescaled matter field Lagrangian is
\be
L_{m} = \frac14 F_{\mu\nu}F^{\mu\nu} + D_\mu\phi^*D^\mu\phi +\partial_\mu\psi \partial^\mu\psi +
m^2 \psi^2|\phi|^2 + \mu^2 (\psi^2 -v^2)^2 \, .
\label{GaugedFLS}
\ee
Here $\cgg$ is the determinant of the space-time metric $g_{\mu\nu}$, $\alpha^2=4\pi G$ is the gravitational coupling with Newton's constant $G$, and $R$ is the Ricci scalar.
In the matter field Lagrangian the $U(1)$ field strength tensor is
$F_{\mu\nu}=\partial_\mu A_\nu-\partial_\nu A_\mu$,
and the covariant derivative of the complex field $\phi$ is $D_\mu\phi = \partial_\mu \phi - ig A_\mu\phi$ with 
%the geometric covariant derivative $\partial_\mu$ and
the gauge coupling constant $g$.

The last two terms represent the symmetry breaking scalar field potential of the model
\be
U(\psi,\phi)=m^2 \psi^2 |\phi|^2 + \mu^2(\psi^2-v^2)^2   \, ,
\label{pot}
\ee
where $m$ and $\mu$ are positive constants.
The global minimum of the potential \re{pot}
corresponds to $\psi=v$, $|\phi|=0$. 
Here the fields assume their vacuum expectation values.
Furthermore, in vacuum $D_\mu \phi =0$, $\partial_\mu \psi =0$, and $F_{\mu\nu}=0$.

With such a choice of the potential, the parameter $\mu$ defines the mass of the excitations of the real scalar component $\psi$,
$m_\psi= \sqrt8 \mu v$. 
The complex scalar $\phi$ becomes massive due to the coupling with the real field $\psi$, $m_\phi = mv$. 
Note that, for any finite values of the parameter $m$,
the complex field becomes massless when the real component is zero. As the real component becomes infinitely heavy, $\mu \to \infty$, it decouples, 
$\psi=1$, and the EMFLS model \re{lag} reduces to the Einstein-Maxwell-Klein-Gordon theory. 
Therefore, one might expect that, for relatively large values of the parameter $\mu$, solutions of the EMFLS theory are similar to the mini boson stars in the gauged EKG model.      
%Without loss of generality, we set $v=1$, $\mu^2=0.25$ and $m=1$.
%

The gauge field acquires a mass due to the coupling with the scalar field $\phi$.
The mass of the gauge excitations $A_\mu$ turns to be zero as $|\phi|=0$.
In this limit the gauge field becomes long-ranged.
Scaling relations allow to set $v=1$ and $m=1$,
leaving us with the parameters $\mu$ and $\alpha$
\cite{Kunz:2019sgn}.

The model \re{lag} is invariant with respect to local $U(1)$ gauge transformations
\be
\phi \to \phi e^{ig\xi(x)}\, ;\quad A_\mu \to A_\mu + \partial_\mu \xi \ .
\label{symmetry}
\ee

The system of the EMFLS field equations can be obtained via variation of the action \re{lag} with respect to the metric, the gauge potential and the scalar fields, respectively
\be
\begin{split}
R_{\mu\nu} -\frac12 R g_{\mu\nu} &= 8\pi G\left( T_{\mu\nu}^{Em} + T_{\mu\nu}^{sc}\right)\, ,\\
\partial_\mu(\sqrt{-\cal g} F^{\mu\nu})& = g \sqrt{-\cal g} j^\nu\, , \\
\end{split}
\label{field-eqs-em}
\ee
where
\be
j_\nu = i( D_\nu \phi^* \, \phi - \phi^* D_\nu \phi )
\label{current}
\ee
is the conserved Noether current associated with the  local $U(1)$ symmetry \re{symmetry},
and the components of the stress-energy tensor
of the electromagnetic and the scalar fields are
\be
\begin{split}
T_{\mu\nu}^{Em} &=F_\mu^\rho F_{\nu\rho} - \frac14 g_{\mu\nu} F_{\rho\sigma}  F^{\rho\sigma}\,,\\
T_{\mu\nu}^{Sc} &=D_\mu\phi^* D_\nu\phi + D_\nu\phi^* D_\mu\phi +  \partial_\mu\psi
\partial_\nu\psi\\
& - g_{\mu\nu}\left(\frac{g^{\rho\sigma}}{2} (D_\rho\phi^* D_\sigma\phi +
D_\sigma\phi^* D_\rho\phi +  \partial_\rho\psi
\partial_\sigma\psi) +  U(\phi,\psi)\right) \, .
\end{split}
\label{Teng}
\ee
The scalar field equations are
\be
\begin{split}
\partial^\mu \partial_\mu\psi & = 2\psi (m^2 |\phi|^2  +2 \mu^2 (1-\psi^2)) \, , \\
D^\mu D_\mu\phi & = m^2 \psi^2 \phi \, .
\end{split}
\label{field-eqs-scalar}
\ee

We are interested in stationary spherically-symmetric solutions of the model \re{lag}.
To construct such solutions numerically we employ a Schwarzschild-like parametrization of the metric
\be
ds^2 = g_{\mu\nu}dx^\mu dx^\nu = -\sigma^2(r)N(r) dt^2 + \frac{dr^2}{N(r)} + r^2\left(d\theta^2 + \sin^2\theta d\varphi^2
\right)
\label{metric}
\ee
with $N(r)=1-\frac{2M(r)}{r}$.
The corresponding parametrization of the scalar fields is
\be
\label{scalans}
\psi=X(r)\, , \qquad  \phi=Y(r)e^{i\omega t}\, ,
\ee
where $\omega$ is the angular frequency of the complex field $\phi$.
Further, in the static gauge the gauge potential can be written as
\be
\label{Aans}
A_{\mu} dx^{\mu} =A_0(r)dt \, .
\ee

The full system of the field equations \re{field-eqs-em}-\re{field-eqs-scalar} can be solved numerically using the
parametrization \re{metric}-\re{Aans}, where we impose
the following set of the boundary conditions:
\begin{itemize}
    \item at $r=0:\qquad \partial_r X = \partial_r Y = \partial_r A_0 = \partial_r N(r)=\partial_r \sigma(r) =0$ \ ,
    \item at $r=\infty:\qquad X=1,\quad Y=0,\quad A_0=0,\quad N(r)=\sigma(r)=1$ \ .
\end{itemize}
As usual, they follow from conditions of regularity of the fields at the origin, from the definition of the vacuum reached at spatial infinity, and from the asymptotic flatness of the metric. 

In the following we present our numerical results, where we first discuss gauged Q-balls in the FLS model in flat space and the turn to the gauged boson stars in the EMFLS model.

\section{Gauged Q-balls} % and boson stars}

%%%%%%%%%%%%%%%%%%%%%%%%%%%%%%%%%%%%%%%%%%%%%%%%%%%%%%%%%%%%%%%%%%%%%%
\begin{figure}[t!]
\begin{center}
\includegraphics[height=.33\textheight,  angle =-90]{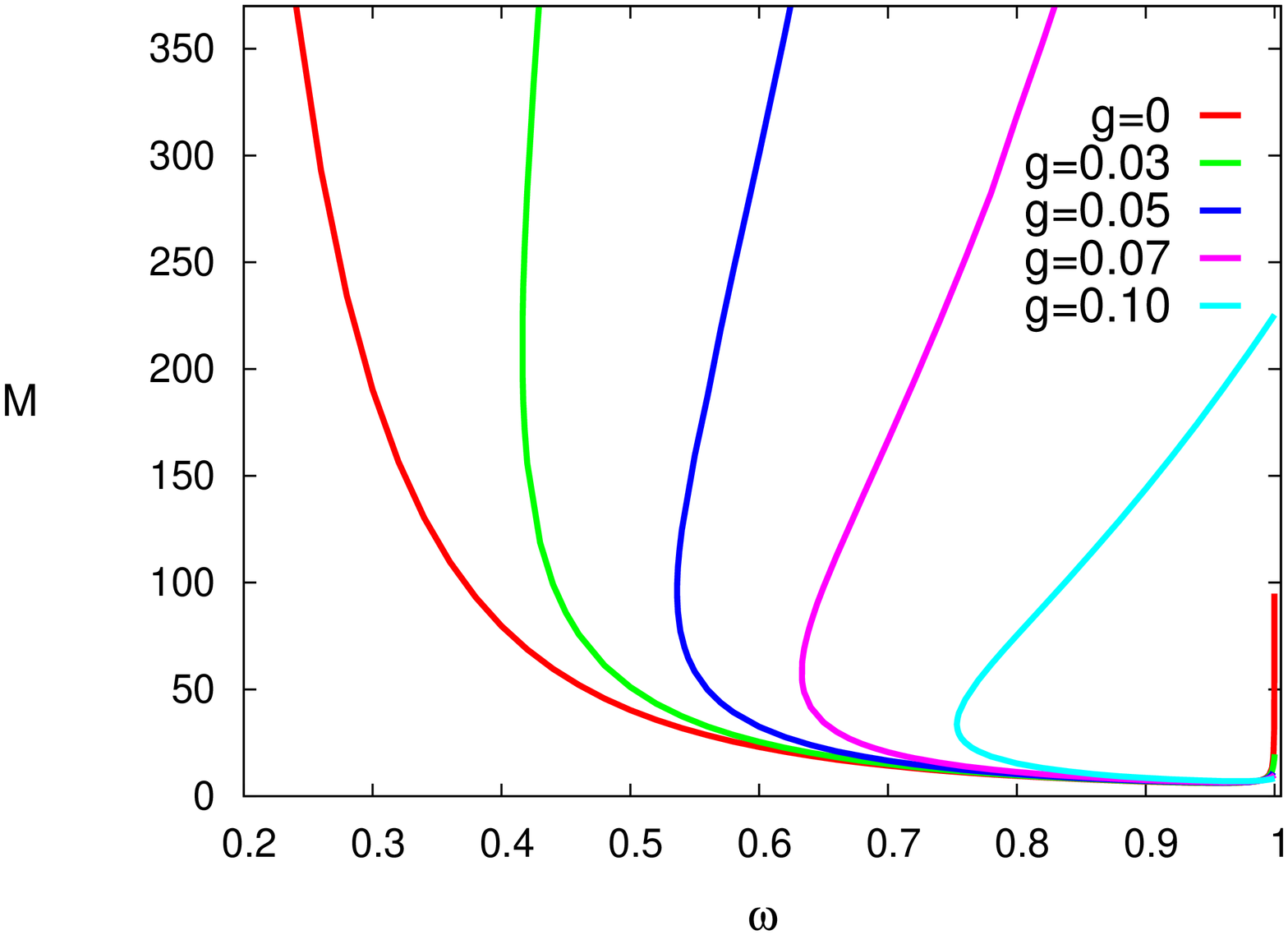}
\includegraphics[height=.33\textheight,  angle =-90]{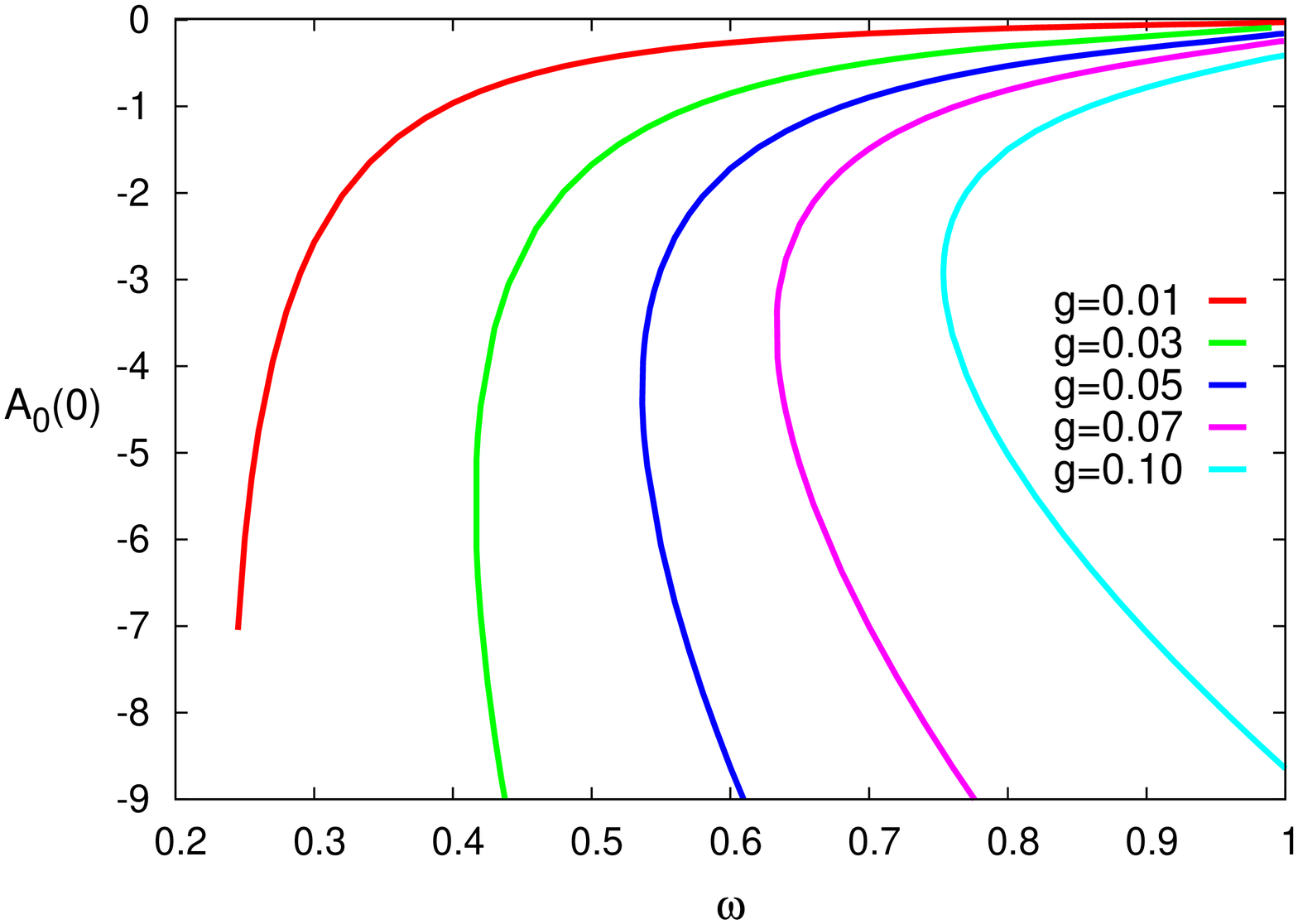}
\includegraphics[height=.33\textheight,  angle =-90]{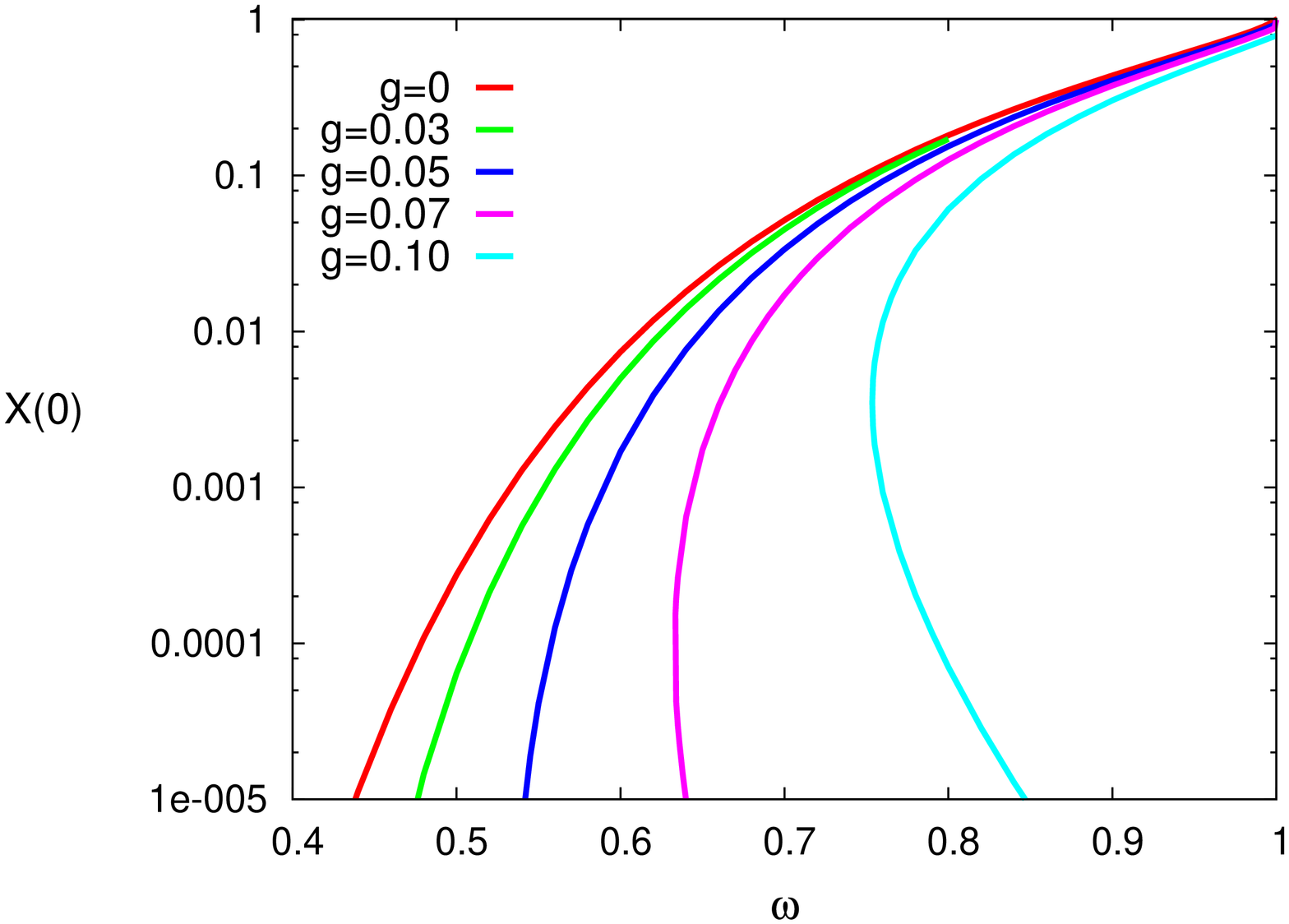}
\includegraphics[height=.33\textheight,  angle =-90]{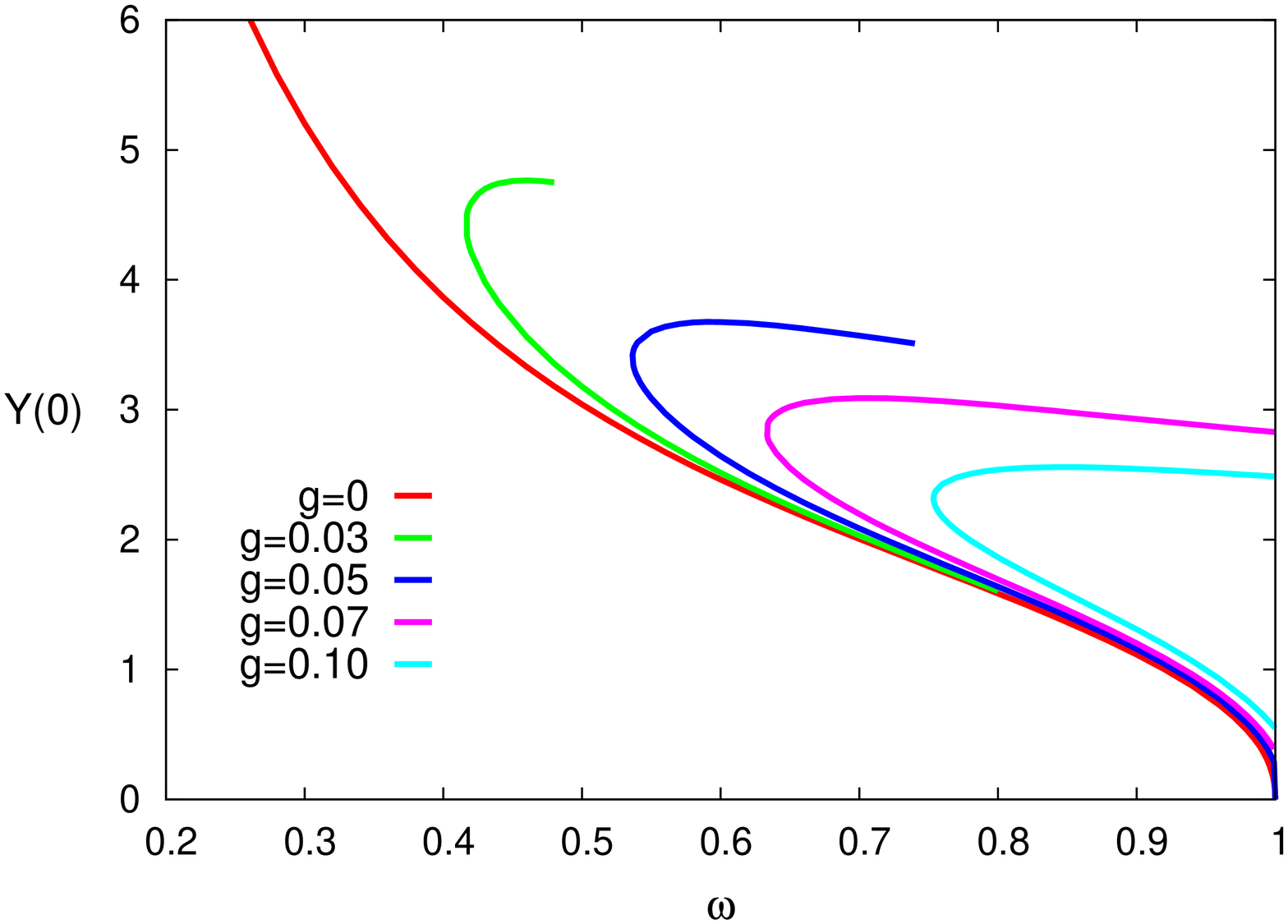}
\end{center}
\caption{\small Gauged FLS Q-balls in Minkowski spacetime.
The total energy of the configuration in units of $8 \pi$ (upper left plot), the values of the gauge potential $A_0$ (upper right plot) and the scalar profile functions $X$ and $Y$ at $r=0$ (lower plots) are displayed as functions
of the angular frequency $\omega$ for $\mu=0.25$, $m=1$ and a set of values of the gauge coupling $g$.}
    \lbfig{fig1}
\end{figure}

We begin by recalling some basic properties of Q-balls in Minkowski spacetime.
Ungauged FLS Q-balls exist for all non-zero values of the frequency $0< \omega < \omega_\mathrm{max}$
\cite{Friedberg:1976me,Levin:2010gp,Loiko:2018mhb},
where the upper bound of the angular frequency $\omega_\mathrm{max}=1$ corresponds to the mass $m=1$ of the complex scalar field. 
This upper bound also holds for the corresponding solutions of the non-renormalizable flat space model with a single complex field and a sextic potential
\cite{Volkov:2002aj,Kleihaus:2005me,Kleihaus:2007vk}.
However, the lower bounds differ for both models,
since the sextic potential leads to a finite minimal value $\omega_\mathrm{min}$ of the frequency, that is determined by the self-interaction of the scalar field.

In both models the energy and the charge of Q-balls typically diverge as the limiting values of the frequency $\omega_\mathrm{min}$ and $\omega_\mathrm{max}$ are approached.
Toward the upper limit, the configurations approach an unbound system of free bosons, whereas toward the lower limit the configurations become more and more strongly bound.
Consequently, there exists a critical value of the frequency, where both energy and charge assume their minimal value.
Only for the case of Q-balls in the FLS model with vanishing mass parameter $\mu$, the energy and the charge tend to zero for $\omega \to \omega_\mathrm{max}$
\cite{Levin:2010gp,Loiko:2018mhb}.

We now address the influence of the presence of the $U(1)$ field on the Q-balls in the FLS model and consider the dependence of the Q-ball properties on the strength of the gauge coupling $g$, choosing a finite value of the mass parameter $\mu$.
When the angular frequency is decreased below the maximal value of the frequency $\omega_\mathrm{max}=1$
gauged FLS Q-balls arise.
%smoothly arise from the perturbative excitations , see Fig.~\ref{fig1}. 
Notably, both the energy and the charge of the gauged flat space Q-balls remain finite in the limit $\omega \to \omega_\mathrm{max}$
\cite{Gulamov:2015fya,Gulamov:2013cra,Panin:2016ooo,Nugaev:2019vru}.

These gauged Q-balls form a branch of solutions which extends backward as $\omega$ decreases. 
Along this branch the properties of the gauged Q-balls are not very different from the corresponding solitons in the ungauged limit. 
The size of the Q-balls increases as $\omega$ decreases.
However, the angular frequency then approaches a finite minimal value $\omega_\mathrm{min}$, where also the energy and the charge are finite. 
In fact, at the finite $\omega_\mathrm{min}$ the derivative of the energy with respect to the frequency $\omega$ diverges. 
This indicates that a bifurcation occurs at $\omega_\mathrm{min}$, and a second branch of gauged Q-balls is encountered.

When studying the behavior of the scalar fields close to $\omega_\mathrm{min}$, we note that the value of the real scalar $\phi$ component at the center of the Q-ball approaches zero.
This corresponds to the massless limit for the complex component $\psi$, which becomes long-ranged in this region.
Furthermore we note, that the energy of the electrostatic repulsion dominates over the scalar interactions, when the bifurcation with the second higher energy branch is approached.

When $\omega$ is increased again along the second branch the characteristic size of the gauged Q-balls continues to increase.
The strong electrostatic interaction then forms a compact domain with a wall that is separating the vacuum with $\phi=1$ on the exterior and confining the massless complex component $\psi$ in the interior. 
This compact domain is blowing up rapidly as the angular frequency approaches its upper critical value, which corresponds again to $\omega_\mathrm{max}=1$.

%In the flat space limit, the properties of $U(1)$ gauged Q-balls in the FLS model (\ref{GaugedFLS}) are different from the corresponding solutions in the model with a one-component complex scalar field and a sextic potential \cite{Loiko:2019gwk}.
%In both cases, there is an upper bound of the angular frequency $\omega_\mathrm{max}=1$, that corresponds to the mass $m=1$ of the complex scalar field. 
%As $\omega \to \omega_\mathrm{max}$, the size of the Q-ball is decreasing and the configurations smoothly approach linearized perturbations around the vacuum. 

%For the ungauged flat space Q-balls in the FLS model in Minkowski spacetime in that limit both the mass and the charge of the configuration rapidly diverge. 

Thus, in the gauged FLS model the Q-balls experience an additional repulsive interaction arising from the gauge sector, that increases with increasing coupling constant $g$.
The minimal angular frequency $\omega_\mathrm{min} \ne 0$ of the gauged FLS Q-balls therefore increases as the gauge coupling increases, and the solutions cease to exist at some maximal critical value of the gauge coupling $g$.
We demonstrate this behavior of the gauged FLS Q-balls in Fig.~\ref{fig1}, where we exhibit the energy $M$ of the configurations for a fixed value of $\mu$ and increasing gauge coupling constant $g$ (upper left), 
as well as the values of the gauge field (upper right) and scalar field functions (lower panels) at the origin.
We emphasize that both the energy and the charge of the $U(1)$ gauged Q-balls remain finite at both ends of the allowed range of angular frequencies, and in particular, at $\omega_\mathrm{max}$ for both branches
(see also \cite{Gulamov:2015fya,Gulamov:2013cra,Panin:2016ooo,Nugaev:2019vru}).

When the mass parameter $\mu$ is set to zero, the model \re{GaugedFLS} has a vanishing potential term.
The resulting Q-balls then carry a long range massless real scalar field with a Coulomb-type asymptotic decay \cite{Levin:2010gp,Loiko:2018mhb,Loiko:2019gwk}.
A peculiar feature of these Q-balls is that there is only one branch of solutions which, as noted above, for the ungauged solitons exists for the whole range of values of the angular frequency $\omega \in [0,1]$.
The mass and the charge of these configurations increase monotonically as $\omega$ decreases. 
In this case an increase of the gauge coupling $g$ does not yield a second branch of solutions. 
Instead only the minimal frequency $\omega_\mathrm{min}$ increases so that the allowed frequency range becomes smaller.

\section{Gauged boson stars}

%%%%%%%%%%%%%%%%%%%%%%%%%%%%%%%%%%%%%%%%%%%%%%%%%%%%%%%%%%%%%%%%%%%%%%
\begin{figure}[t!]
\begin{center}
\includegraphics[height=.33\textheight,  angle =-90]{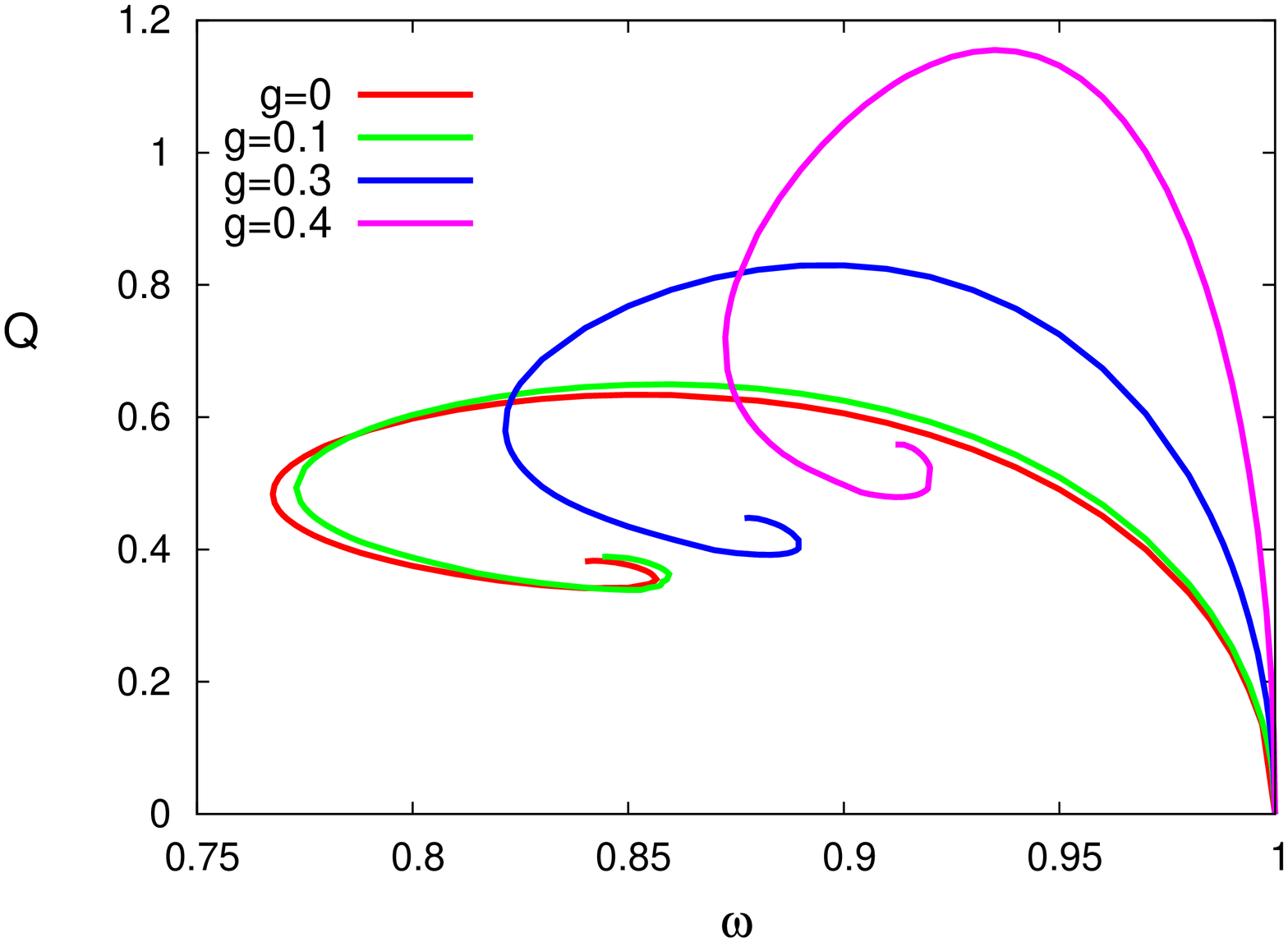}
\includegraphics[height=.33\textheight,  angle =-90]{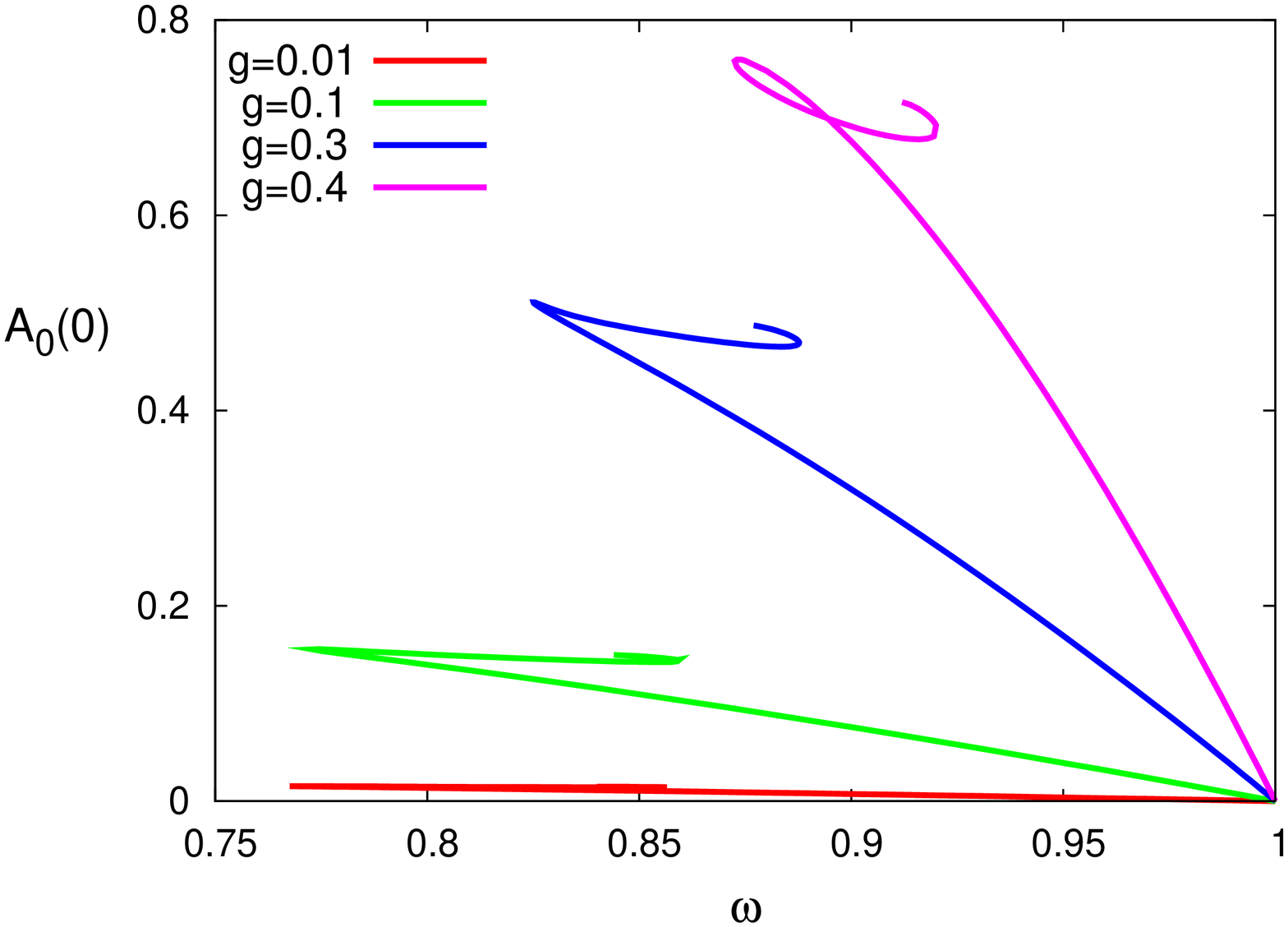}
\includegraphics[height=.33\textheight,  angle =-90]{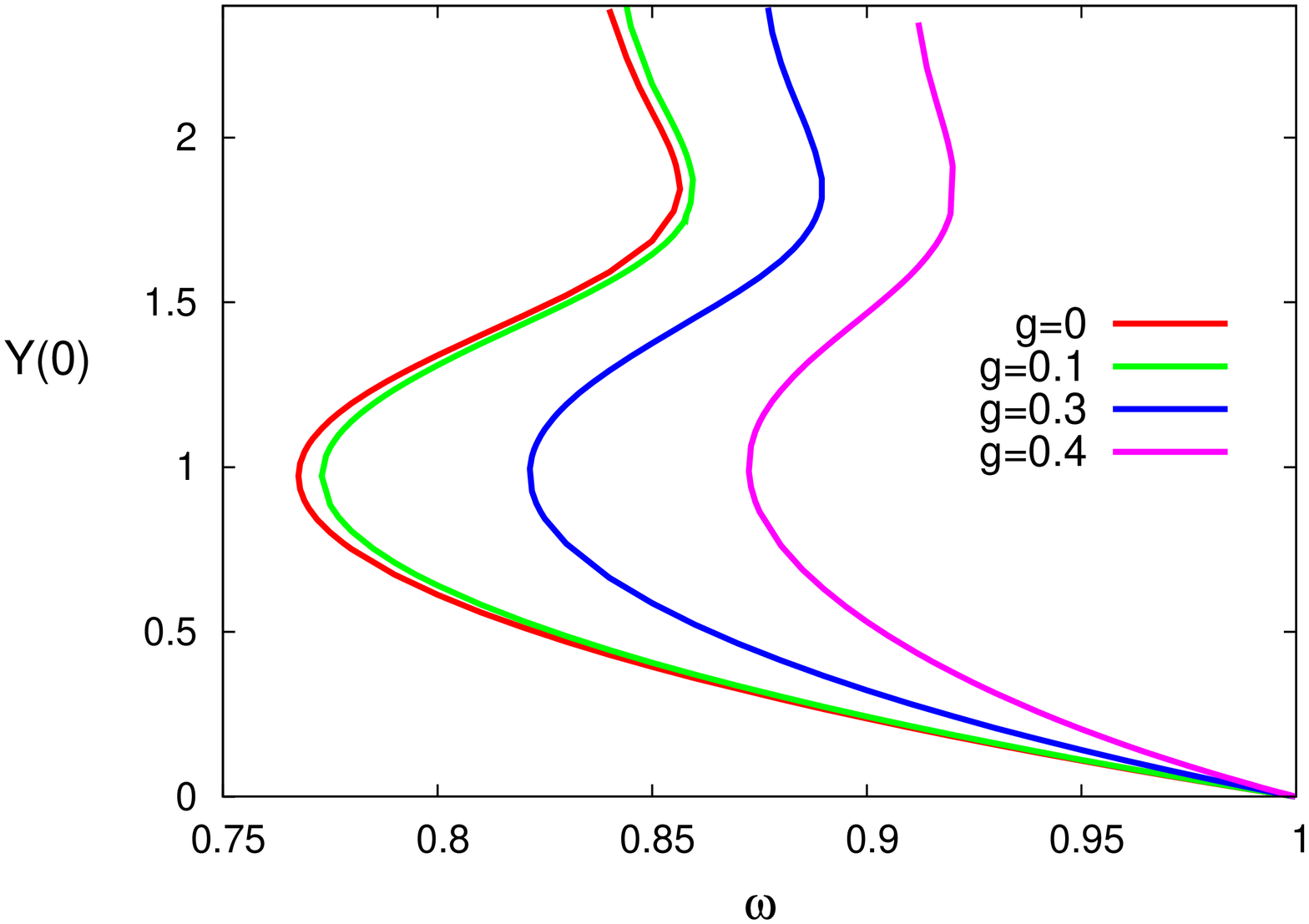}
\includegraphics[height=.33\textheight,  angle =-90]{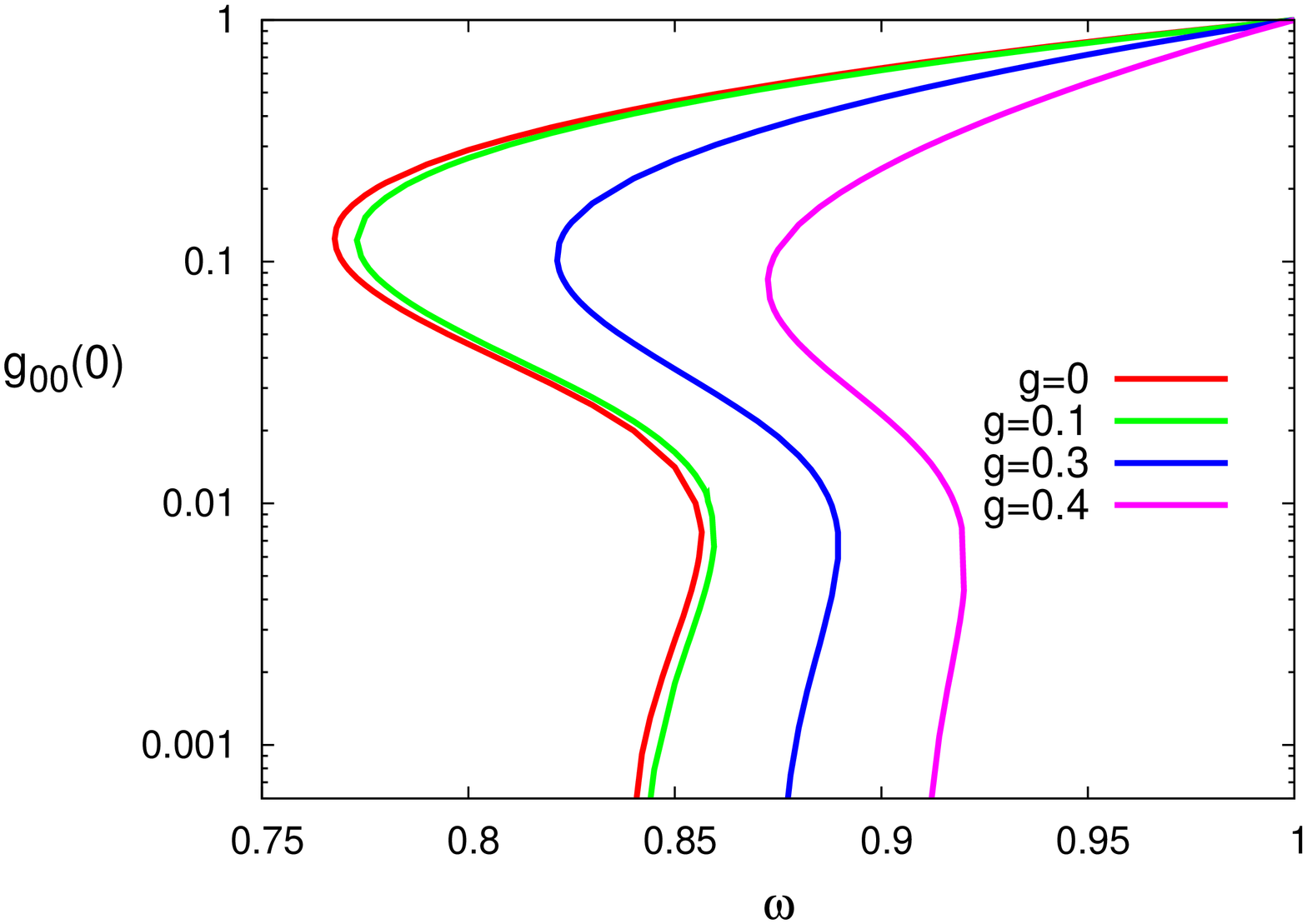}
\end{center}
\caption{\small Gauged EKG boson stars.
The total charge of the gauged boson star in units of $8 \pi$ (upper left plot), the values of the gauge potential $A_0$ (upper right plot), the scalar profile function $Y$ and the metric component $g_{00}$ at $r=0$ (lower plots) are displayed as functions of the angular frequency $\omega$ for $m=1$ for a set of values of the gauge coupling $g$.}
    \lbfig{fig2}
\end{figure}

When gravity is coupled to Q-balls boson stars arise.
Now the additional attractive interaction changes the pattern observed in flat space. 
Boson stars arise smoothly from the vacuum when the angular frequency $\omega$ is decreased below its maximal value $\omega_\mathrm{max}=1$, that corresponds to the mass of the complex scalar field as in flat space.
However, in this limit the mass and charge of the boson stars go to zero.
With decreasing $\omega$ the mass and charge of the boson stars increase, until a maximum is reached. 
For mini-boson stars with a mass term only this corresponds to the global maximum, whereas for soliton boson stars with a sextic potential the global maximum is only reached beyond a local minimum.
But in both cases the mass and charge of the boson stars enter a spiraling phase after the global maximum
\cite{Friedberg:1986tp,Friedberg:1986tq,Kleihaus:2005me,Kleihaus:2007vk}.

Here we are interested in the additional effect provided by the coupling to a $U(1)$ gauge field.
We therefore start by recalling the properties of gauged boson stars in the simpler Einstein-Klein-Gordon (EKG) model. 
We demonstrate the dependence of the charge on the frequency for gauged EKG boson stars in Fig.~\ref{fig2} (upper left), where we vary the gauge coupling $g$. 
The spiral structure is clearly visible, although we have only included the first few turns of the spirals.
The figure also shows the values at the origin of the functions $A_0$ (upper right) and $Y$ (lower left), and of the metric component $g_{00}$ (lower right). 
The latter two exhibit oscillations as the mass and the charge spiral.

When the gauge coupling $g$ is increased from zero, the additional repulsion leads to larger values of the mass and the charge of the boson stars.
At the same time the minimal frequency $\omega_\mathrm{min}$ increases, thus reducing the frequency interval where gauged boson stars exist.
As noted before \cite{Pugliese:2013gsa}, there is a maximal value of the gauge coupling beyond which no further gauged boson stars exist.

We now turn to gauged boson stars in the FLS model.
Since the scenario for the evolution of gauged boson stars with two long-range fields can be very different
from the evolution with only a massless $U(1)$ field,
%\cite{Loiko:2018mhb,Kunz:2019sgn},
we consider these two cases of finite and vanishing mass parameter $\mu$ subsequently in the following, discussing first the more general case of a finite mass parameter $\mu$.

\subsection{Finite mass parameter
\boldmath {$\mu \ne 0$} \unboldmath}

%%%%%%%%%%%%%%%%%%%%%%%%%%%%%%%%%%%%%%%%%%%%%%%%%%%%%%%%%%%%%%%%%%%%%%
\begin{figure}[t!]
\begin{center}
\includegraphics[height=.33\textheight,  angle =-90]{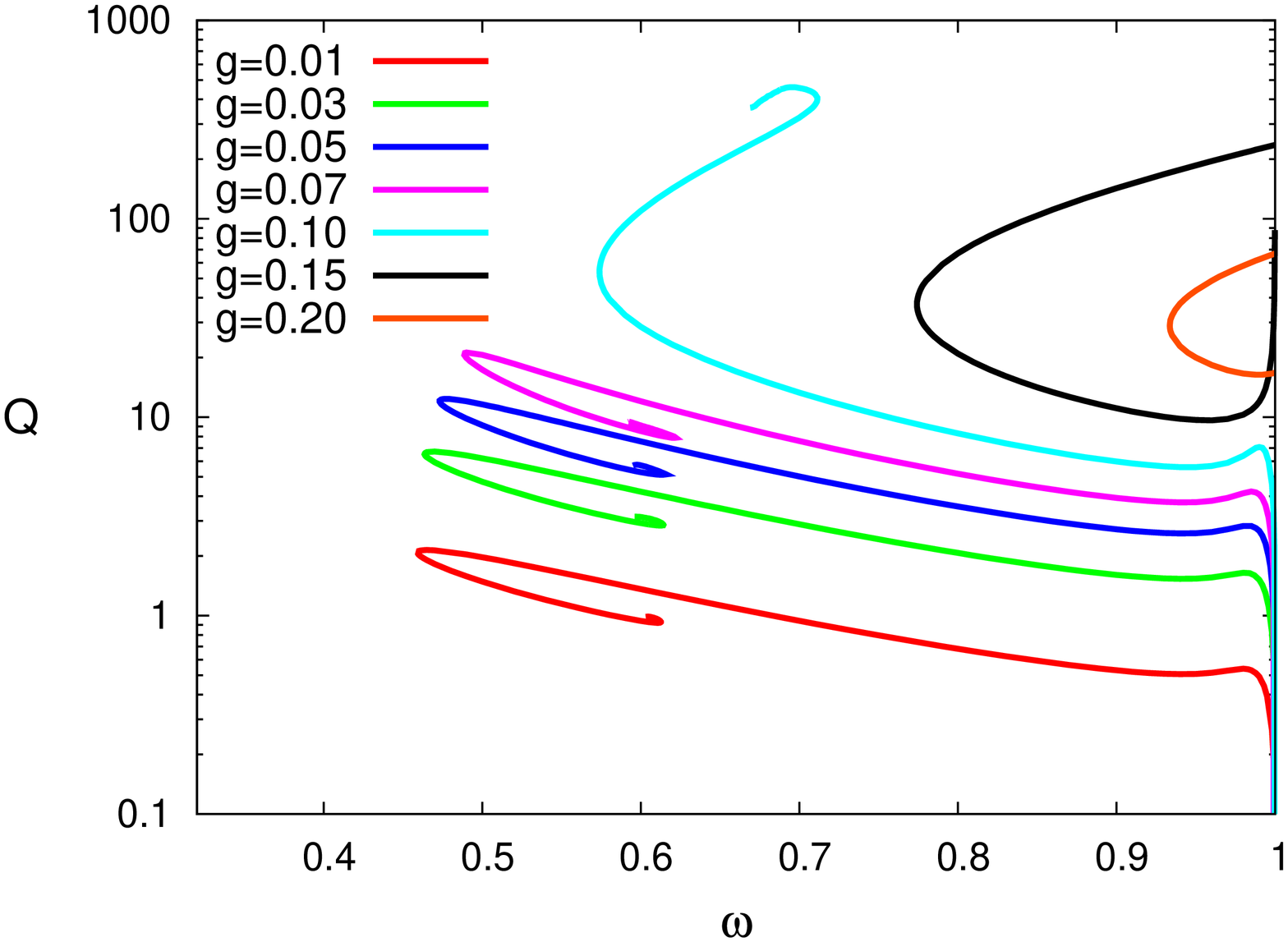}
\includegraphics[height=.33\textheight,  angle =-90]{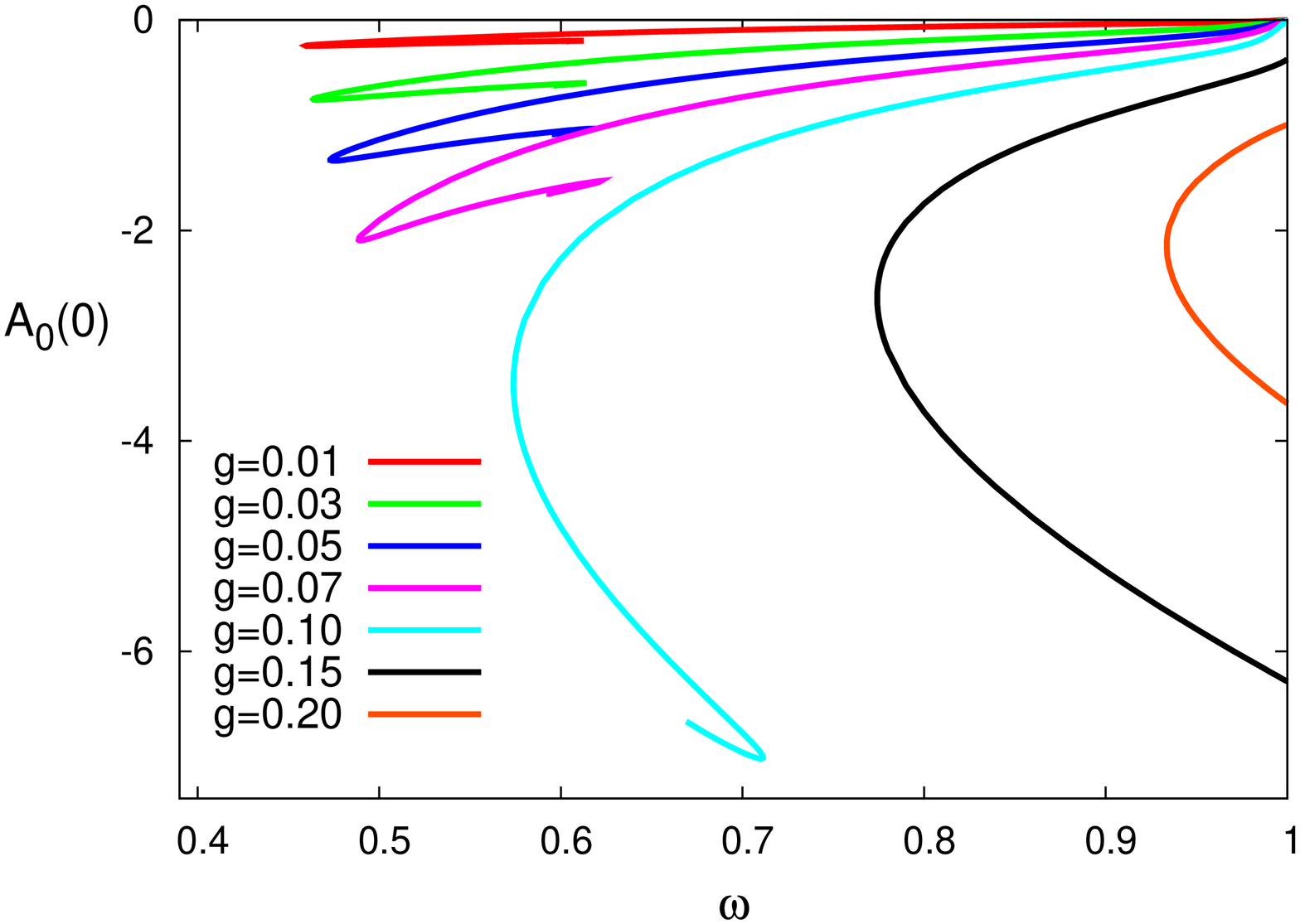}
\includegraphics[height=.33\textheight,  angle =-90]{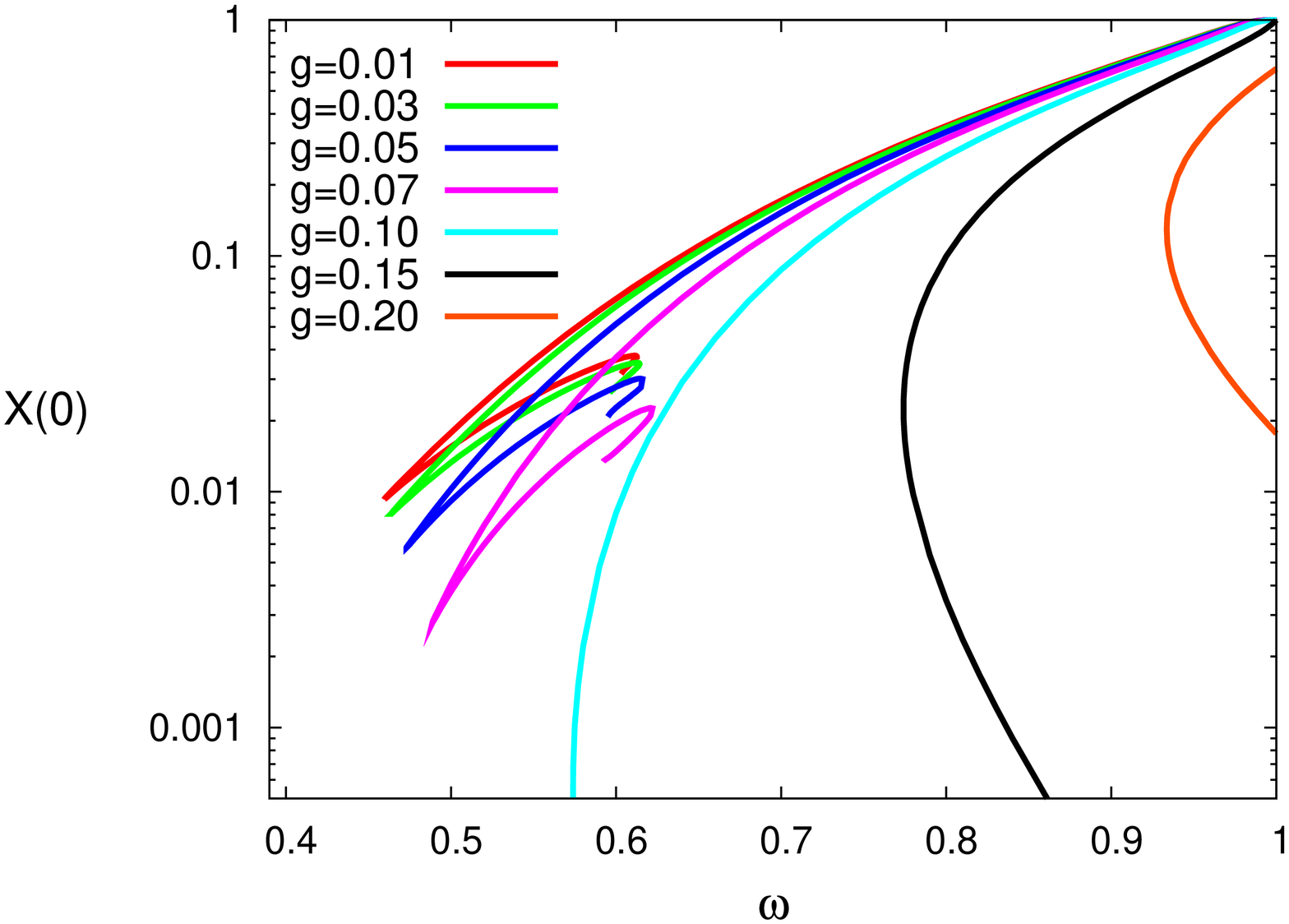}
\includegraphics[height=.33\textheight,  angle =-90]{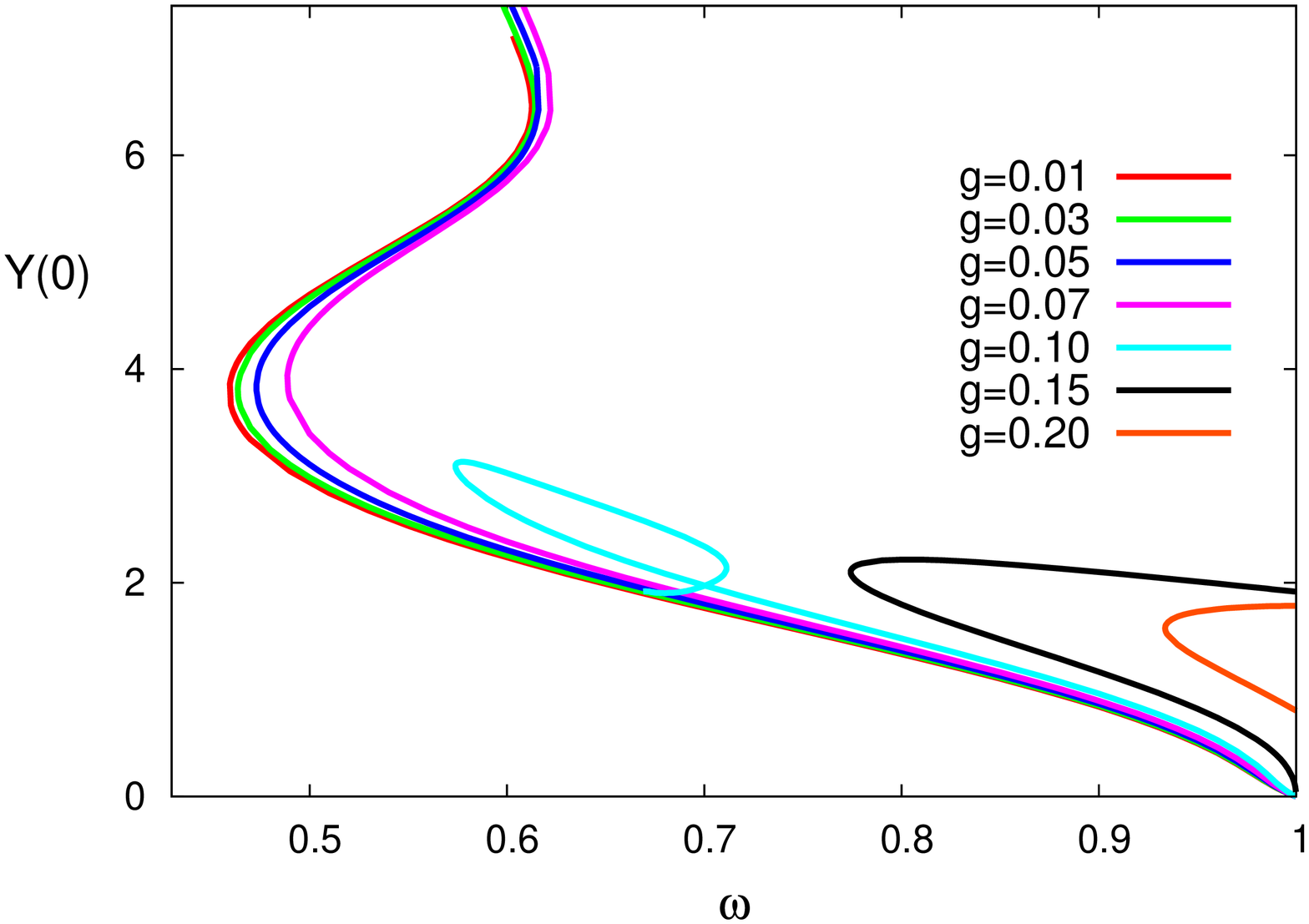}
\includegraphics[height=.33\textheight,  angle =-90]{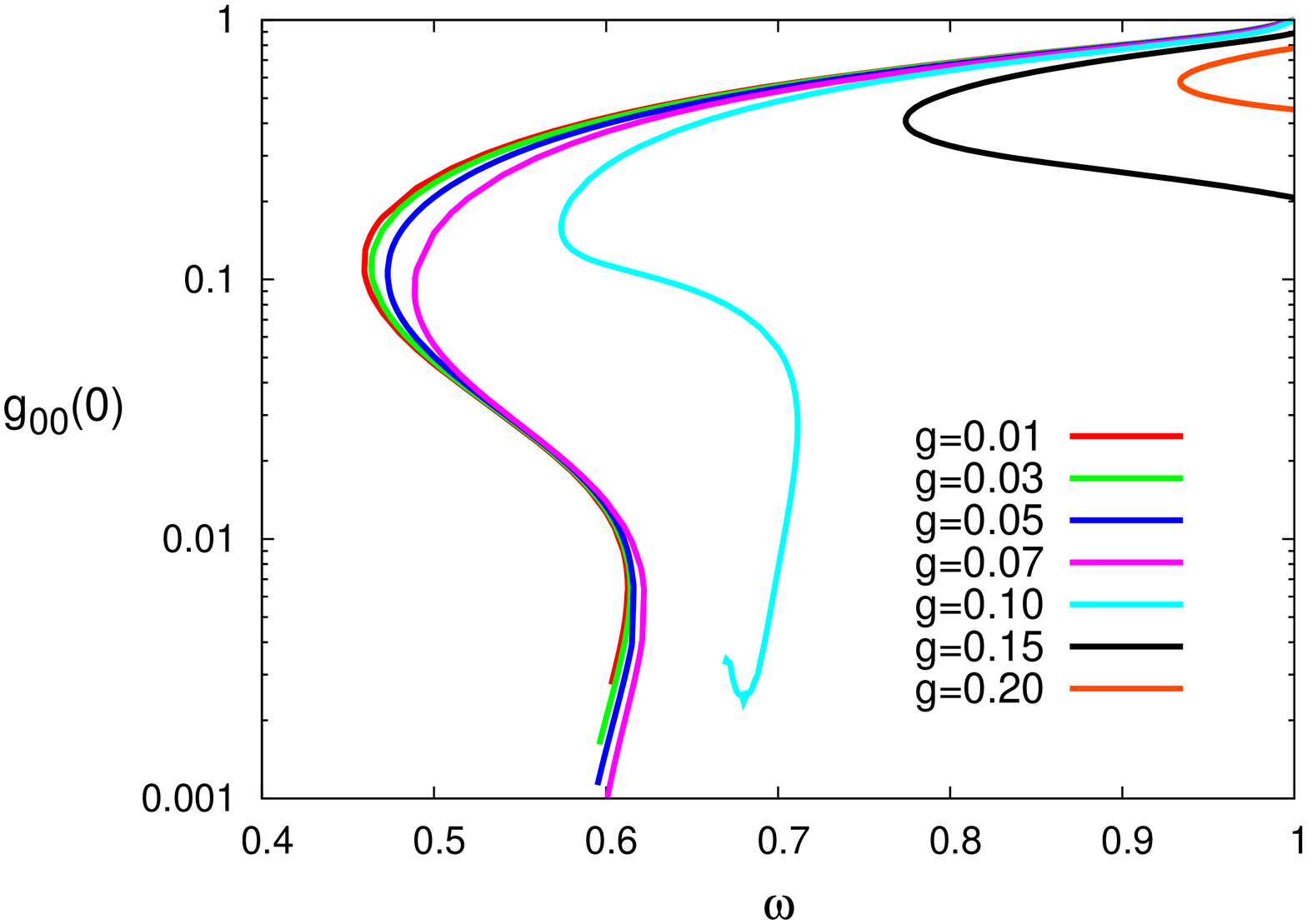}
\end{center}
\caption{\small Gauged Einstein-FLS boson stars.
The total charge of the solutions (upper left plot), the values of the gauge potential $A_0$ (upper right plot), the scalar functions $X$, $Y$ (middle plots), and the metric component $g_{00}$ at $r=0$ (lower plot) are displayed as functions of the angular frequency $\omega$ for $\alpha=0.3$ for a set of values of the gauge coupling $g$.}
    \lbfig{fig3}
\end{figure}

%%%%%%%%%%%%%%%%%%%%%%%%%%%%%%%%%%%%%%%%%%%%%%%%%%%%%%%%%%%%%%%%%%%%%
\begin{figure}[t!]
\begin{center}
\includegraphics[height=.33\textheight,  angle =-90]{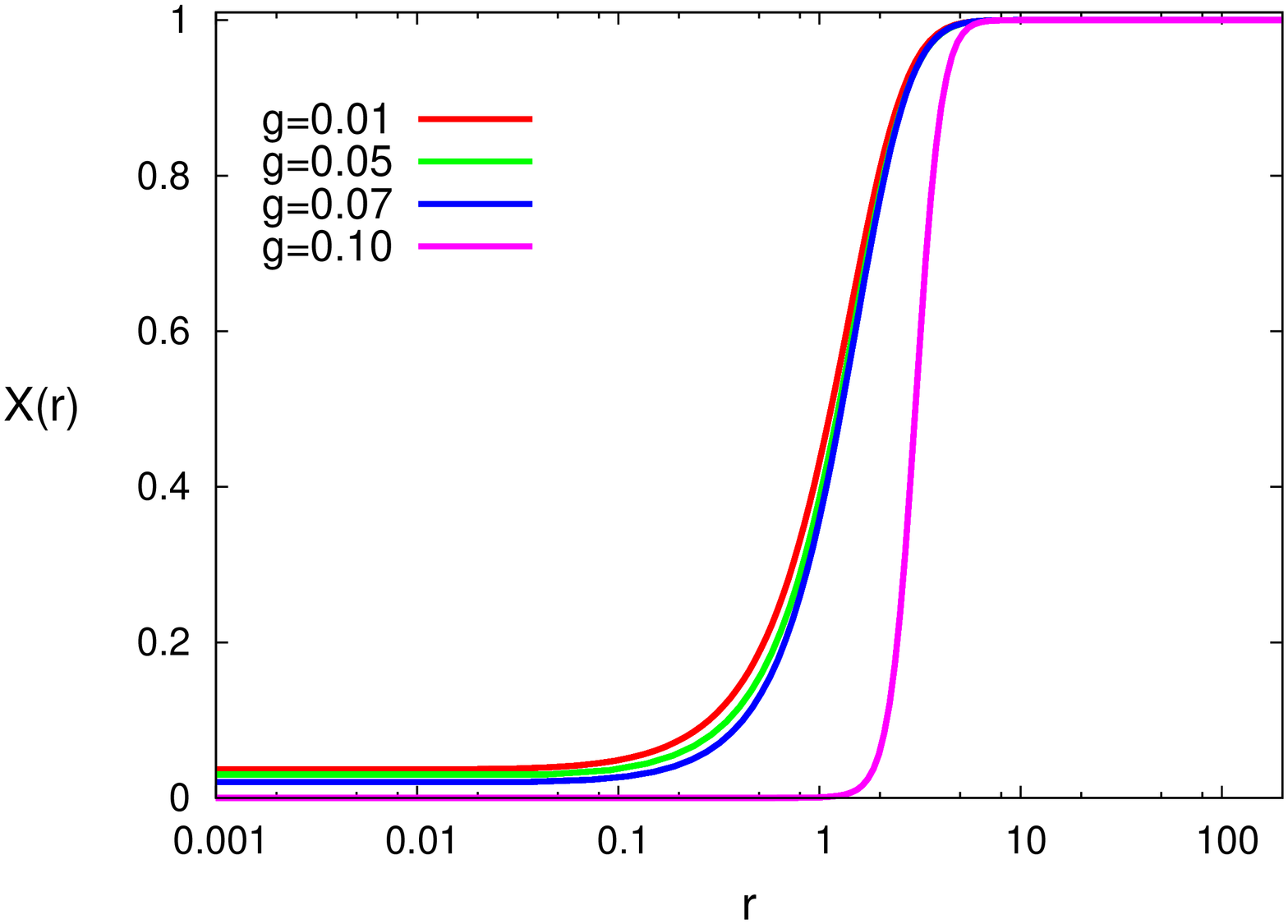}
\includegraphics[height=.33\textheight,  angle =-90]{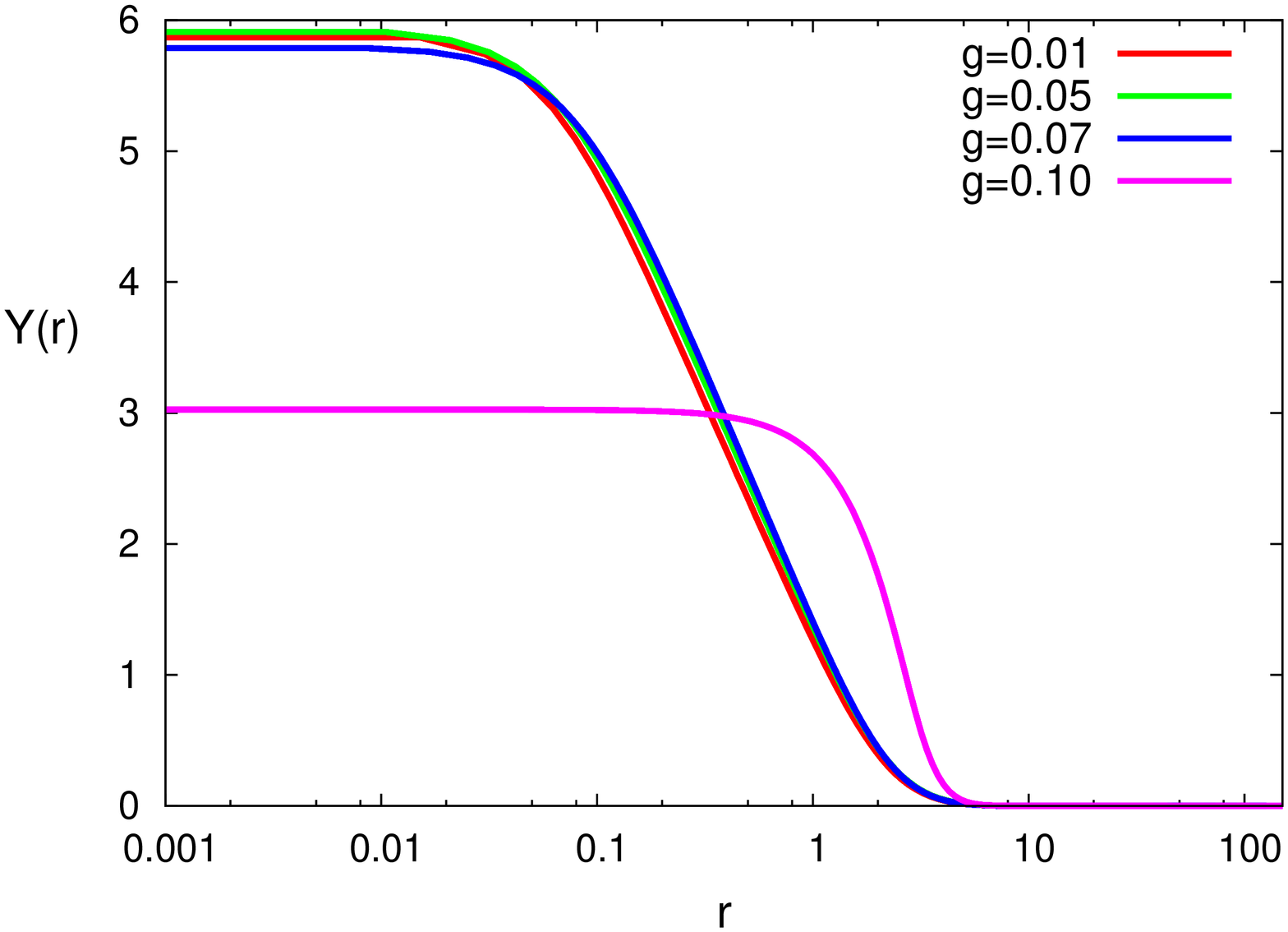}
\includegraphics[height=.33\textheight,  angle =-90]{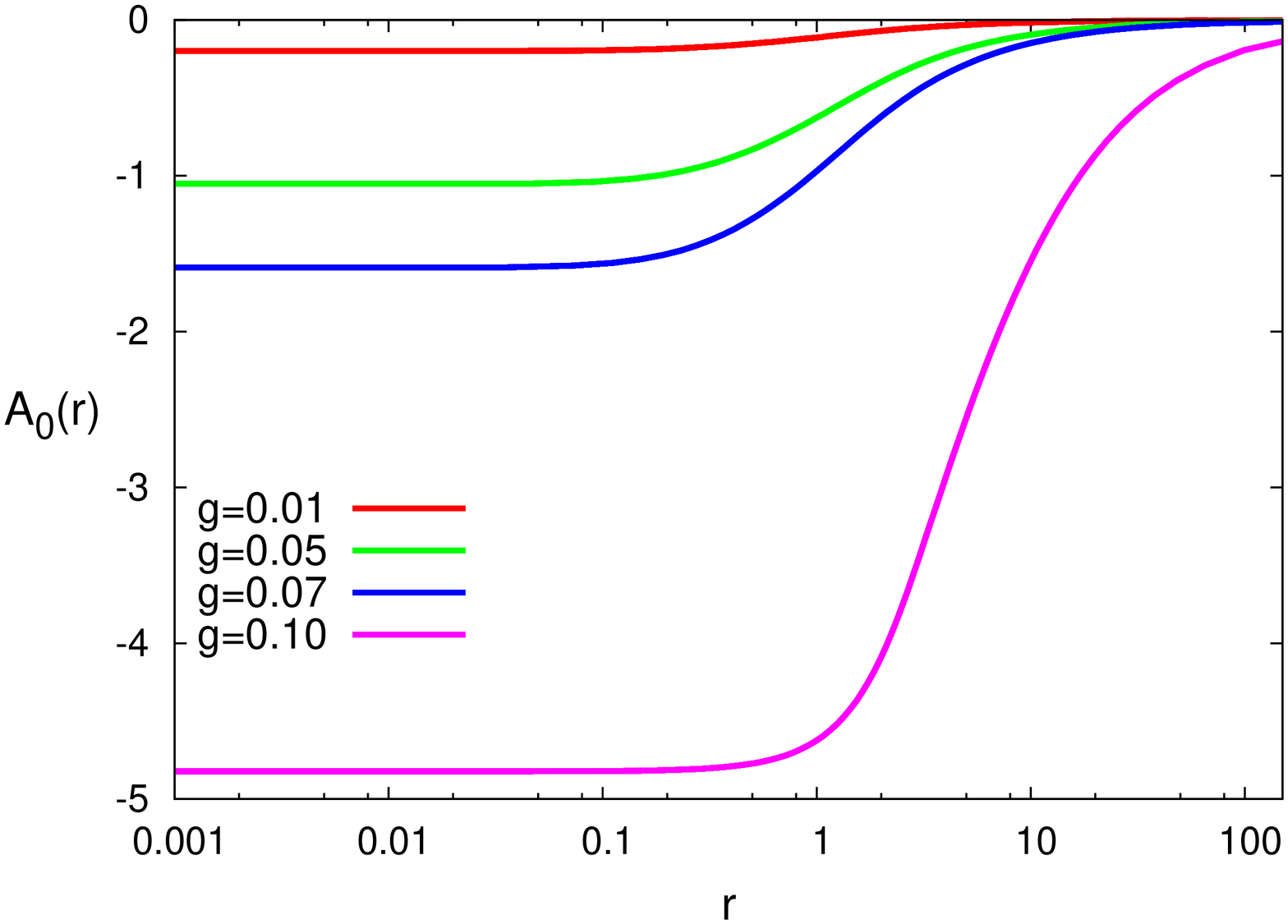}
\includegraphics[height=.33\textheight,  angle =-90]{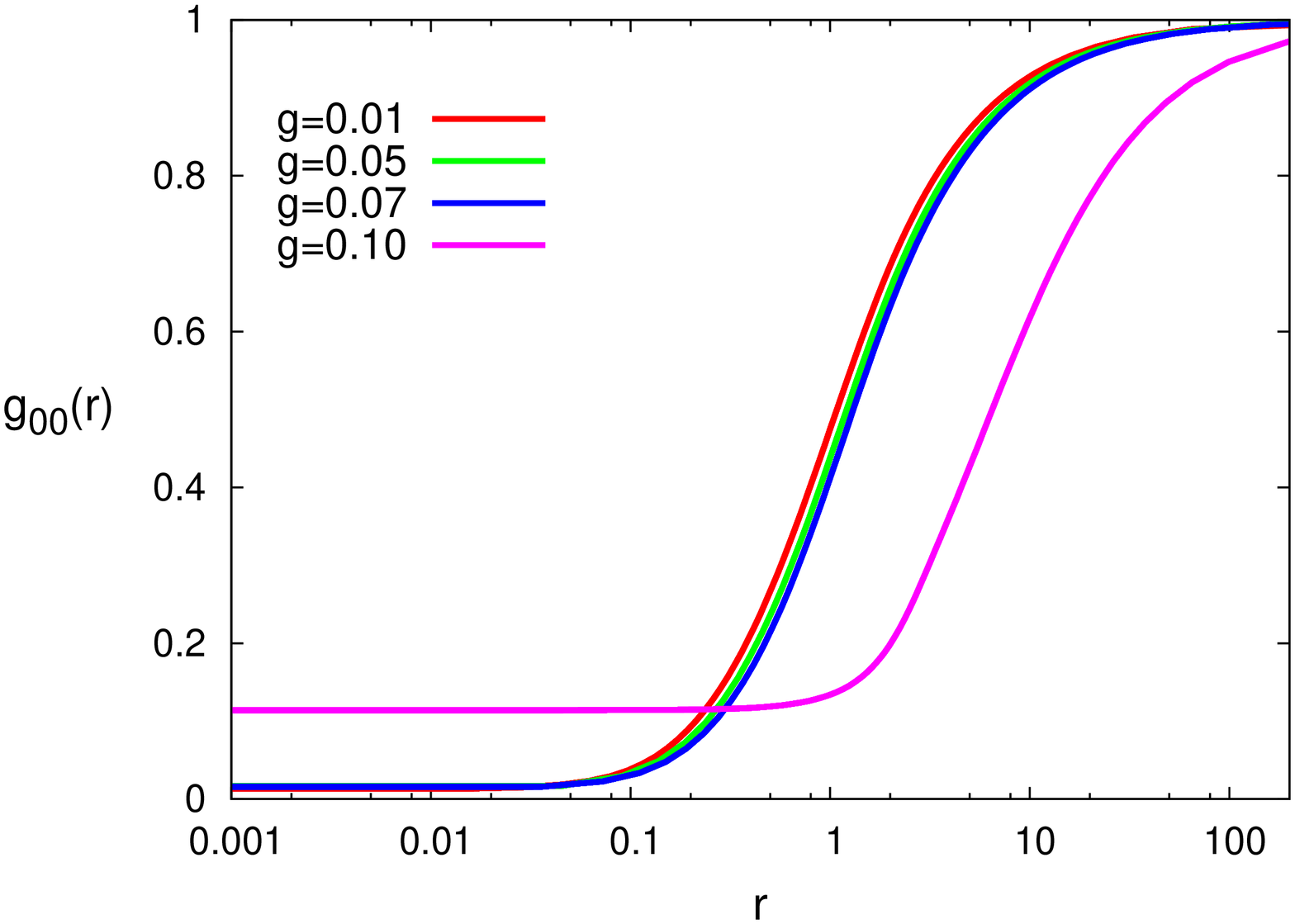}
\end{center}
\caption{\small Compact gauged Einstein-FLS boson stars on the second branch.
The profile functions of the scalar fields $X(r)$ and $Y(r)$ (upper plots), the gauge potential $A_0(r)$ (lower right plot), and the metric component $g_{00}(r)$  (lower left plot) are displayed as functions of the angular frequency $\omega$ for $\alpha=0.3$, $\omega=0.6$ for a set of values of the gauge coupling $g$.}
    \lbfig{fig4}
\end{figure}

%%%%%%%%%%%%%%%%%%%%%%%%%%%%%%%%%%%%%%%%%%%%%%%%%%%%%%%%%%%%%%%%%%%%%
\begin{figure}[t!]
\begin{center}
\includegraphics[height=.33\textheight,  angle =-90]{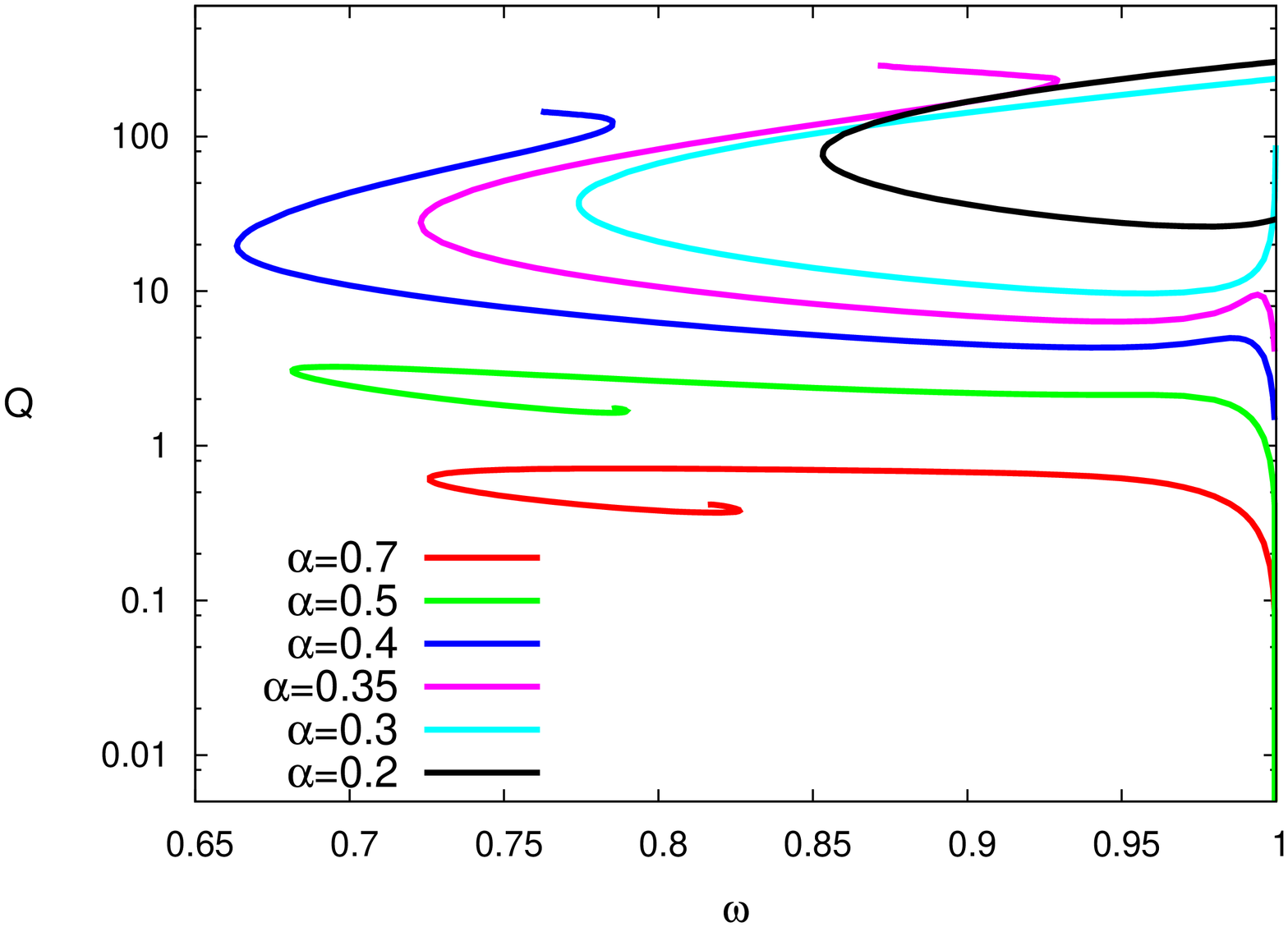}
\includegraphics[height=.33\textheight,  angle =-90]{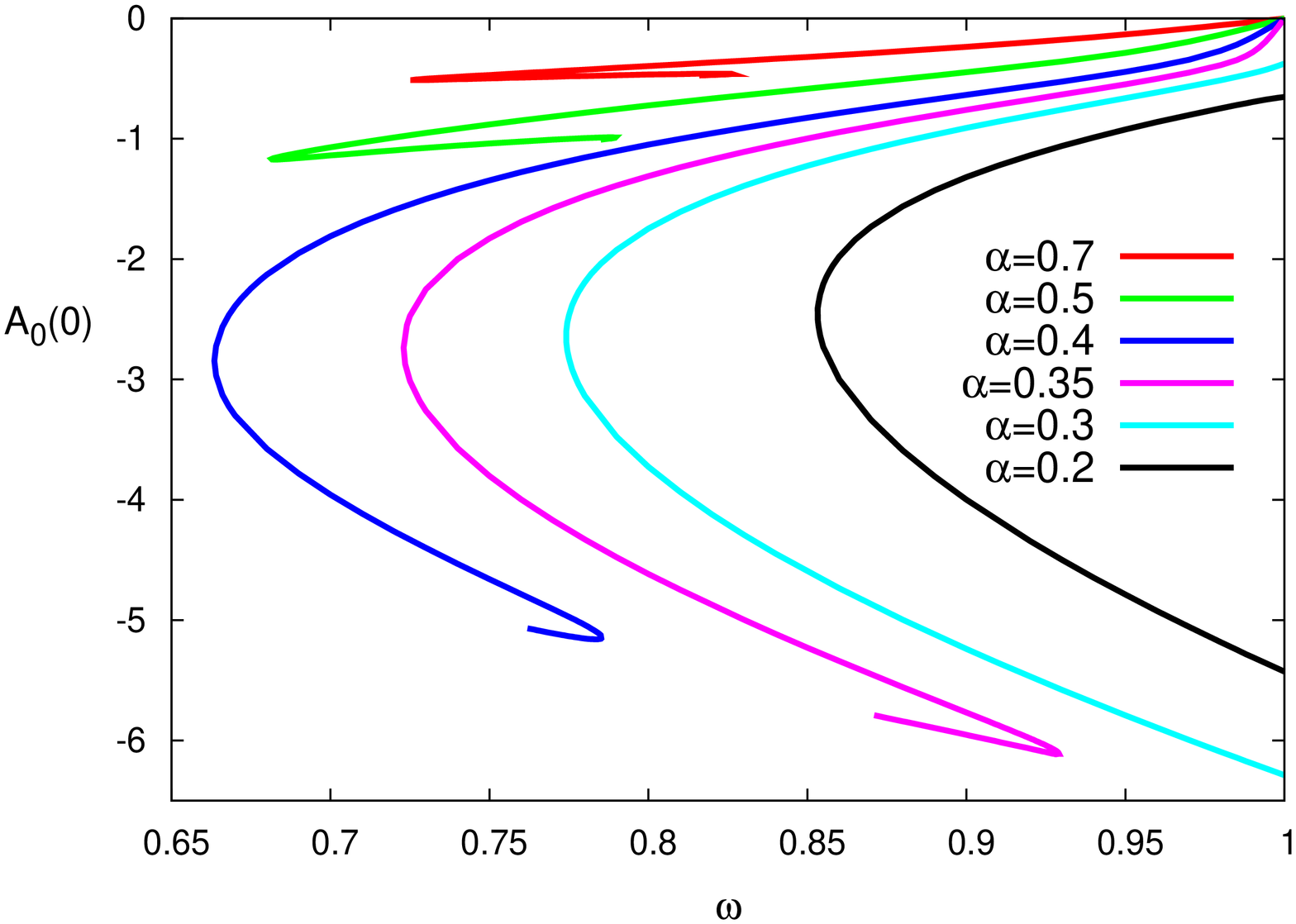}
\includegraphics[height=.33\textheight,  angle =-90]{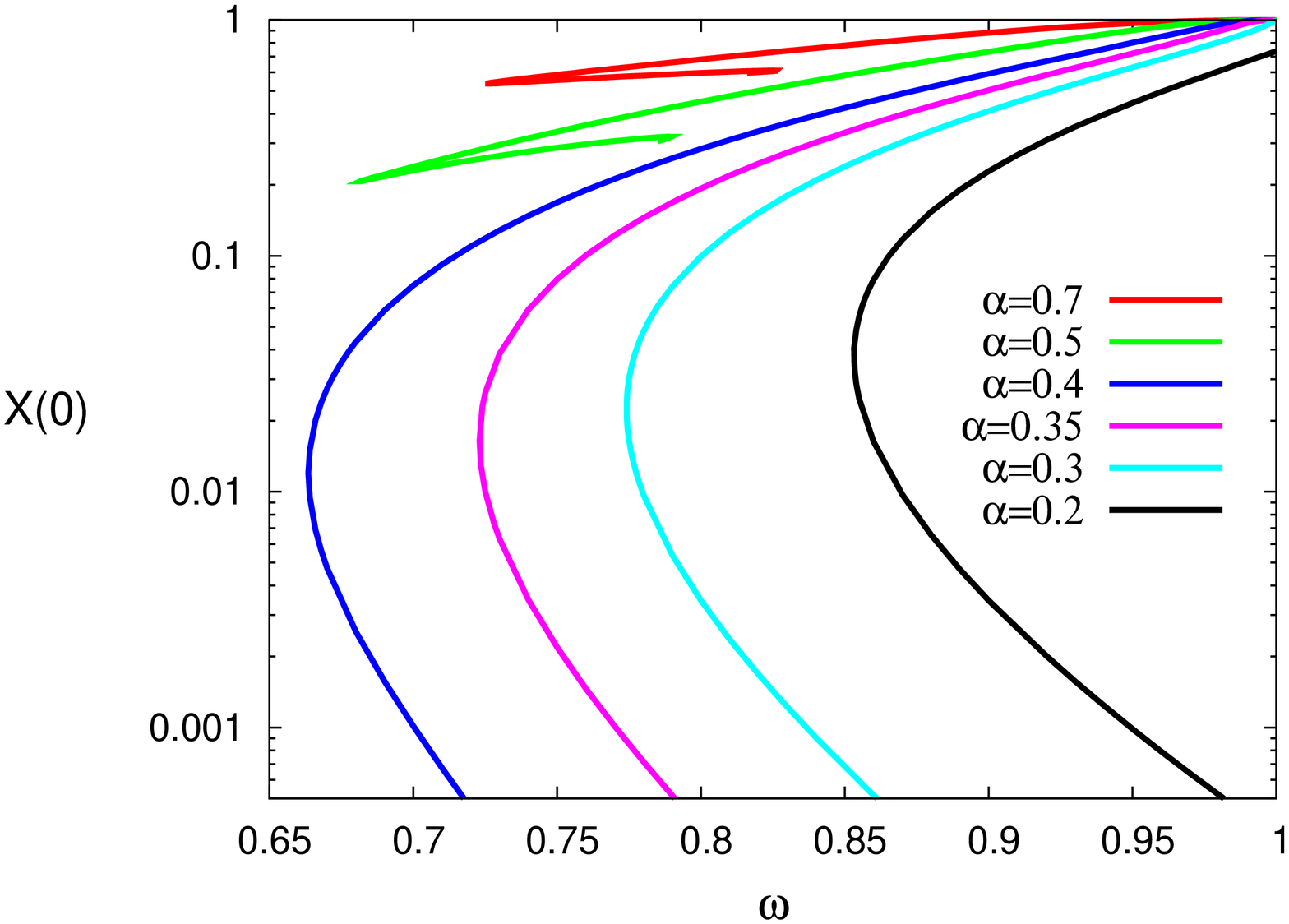}
\includegraphics[height=.33\textheight,  angle =-90]{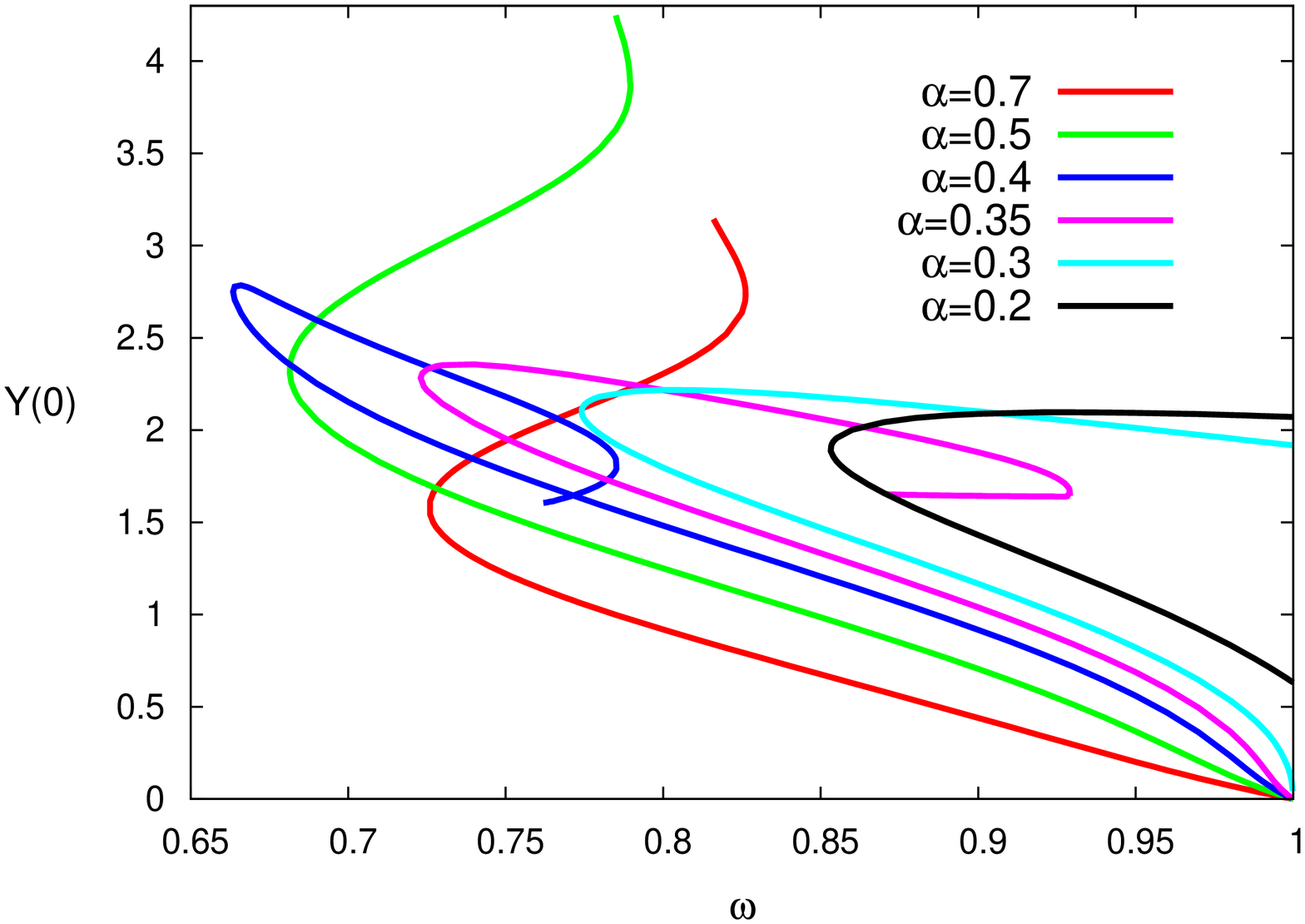}
\includegraphics[height=.33\textheight,  angle =-90]{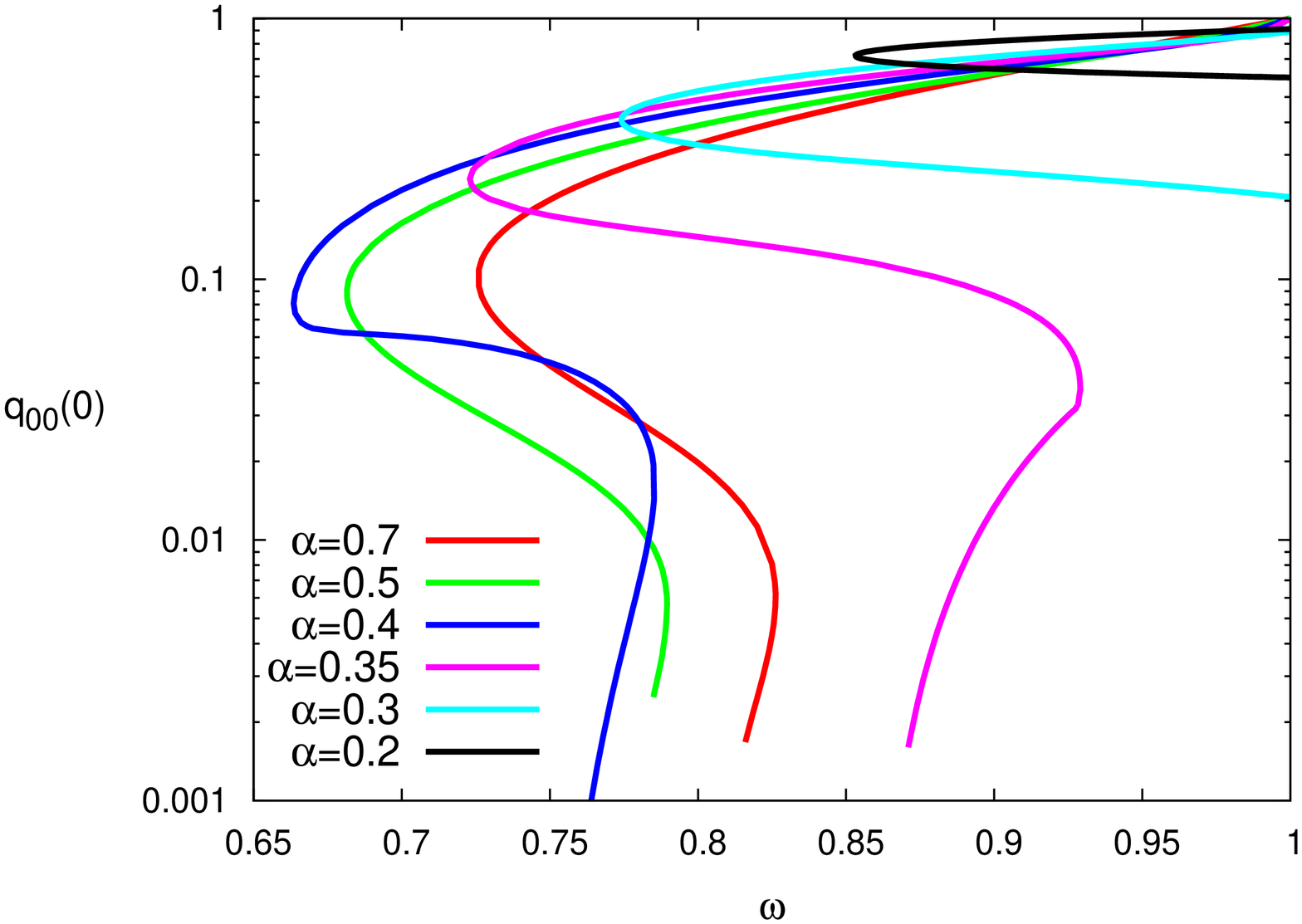}
\end{center}
\caption{\small Gauged Einstein-FLS boson stars.
The total charge of the solutions in units of $8 \pi$ (upper left plot), the values of the gauge potential $A_0$ (upper right plot), the scalar profile functions $X$, $Y$ and the metric component $g_{00}$ at $r=0$ (lower plots) are displayed as functions of the angular frequency $\omega$ for $g=0.15$ for a set of values of the gravitational coupling $\alpha$.}
    \lbfig{fig5}
\end{figure}

%%%%%%%%%%%%%%%%%%%%%%%%%%%%%%%%%%%%%%%%%%%%%%%%%%%%%%%%%%%%%%%%%%%%
\begin{figure}[t!]
\begin{center}
\includegraphics[height=.33\textheight,  angle =-90]{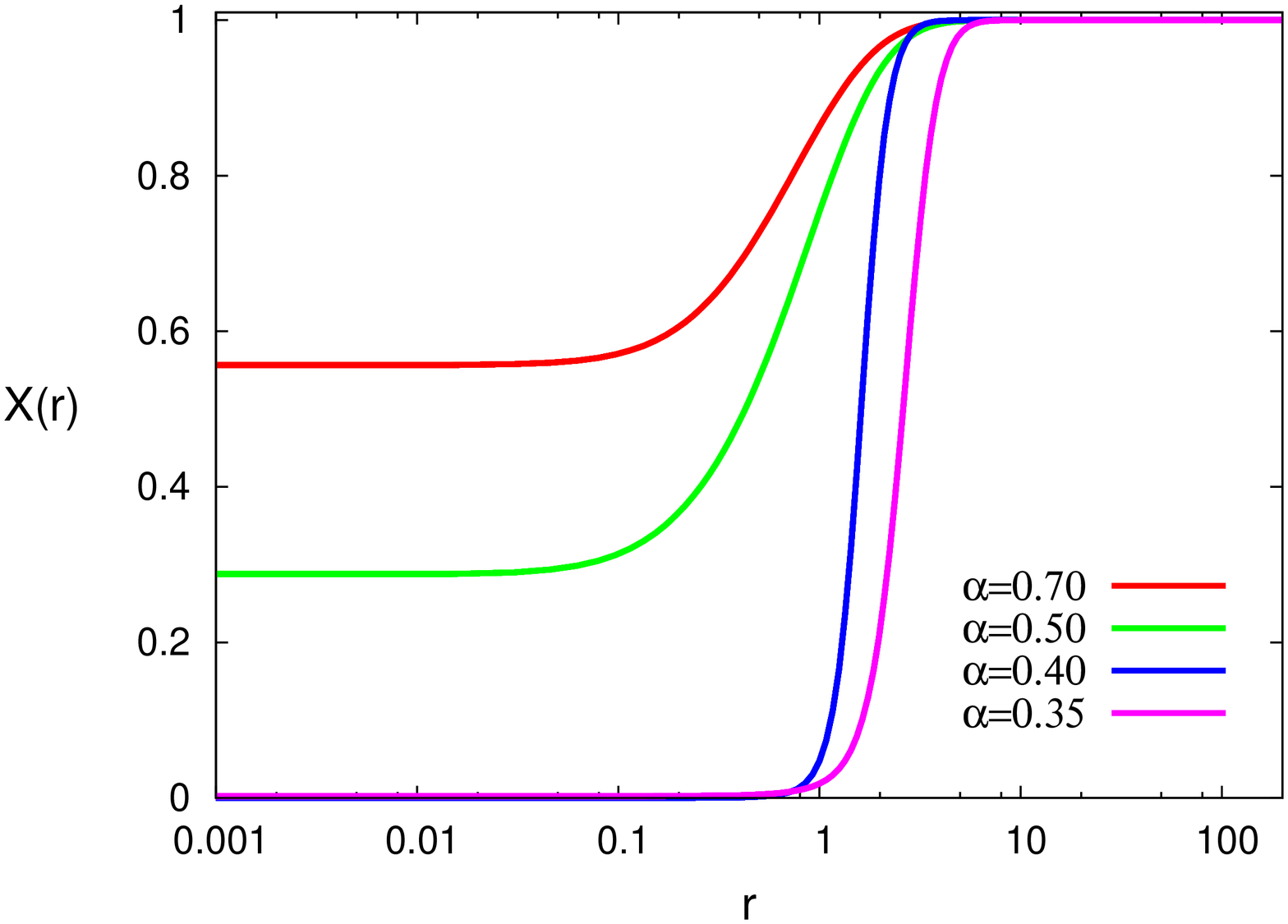}
\includegraphics[height=.33\textheight,  angle =-90]{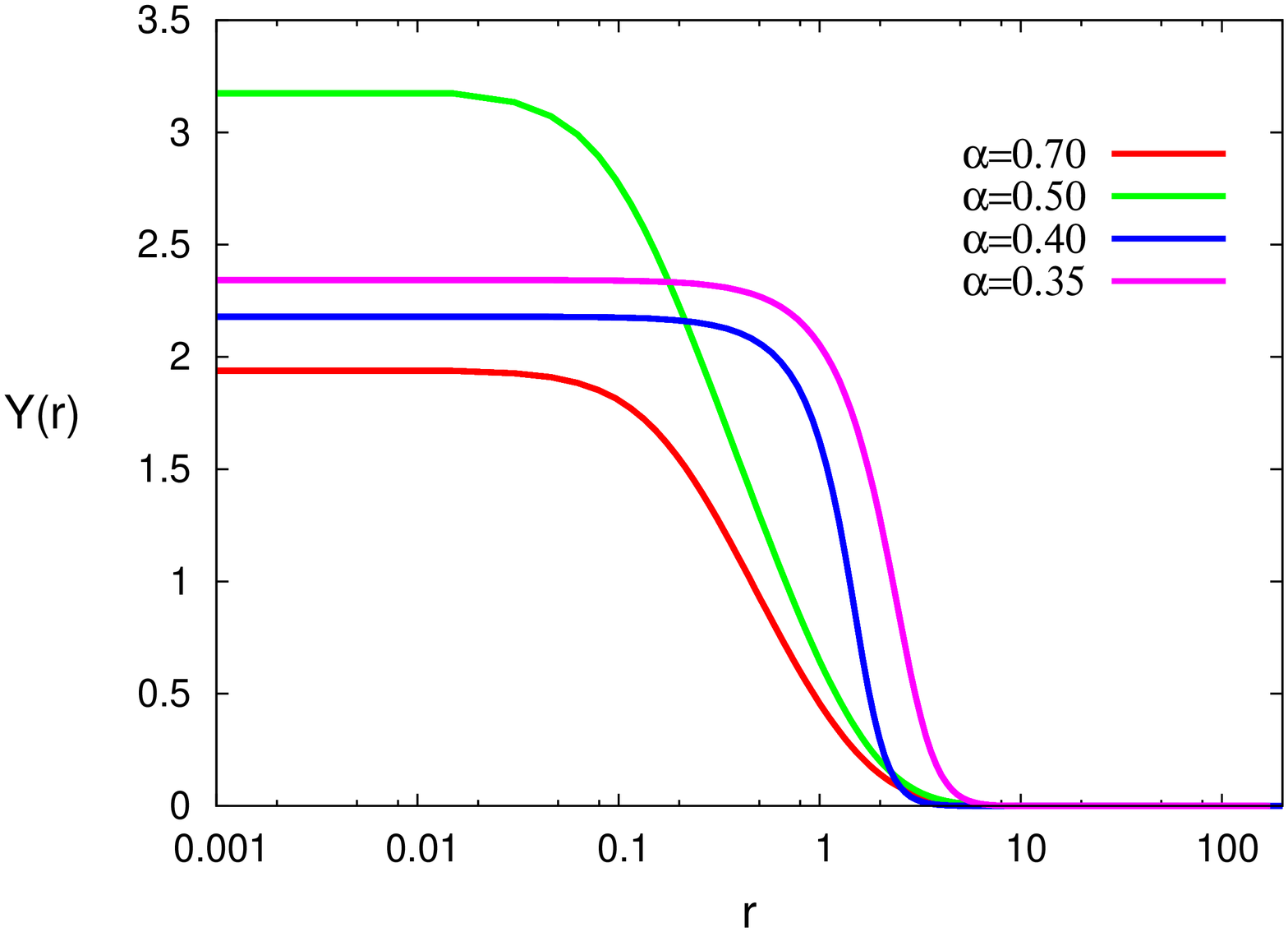}
\includegraphics[height=.33\textheight,  angle =-90]{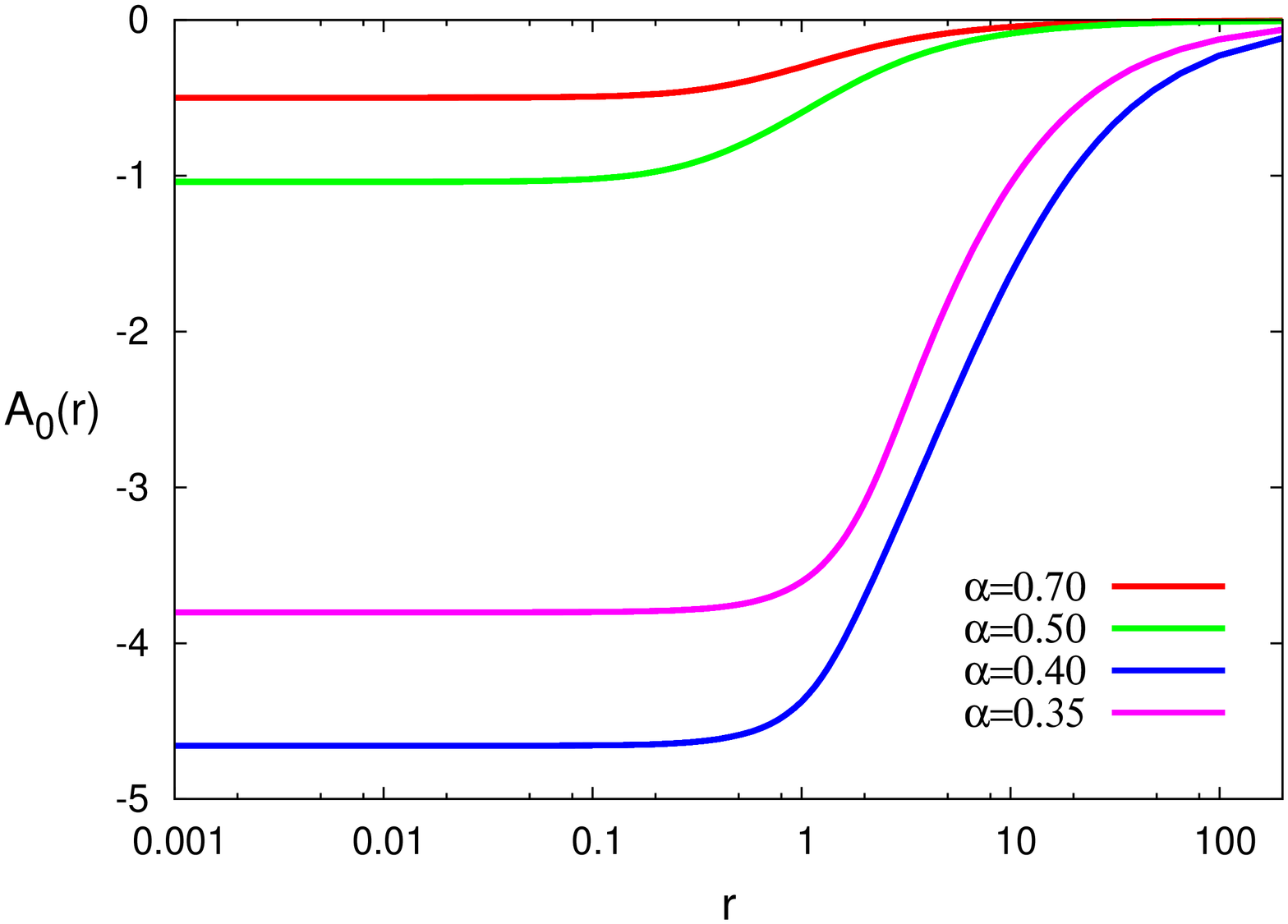}
\includegraphics[height=.33\textheight,  angle =-90]{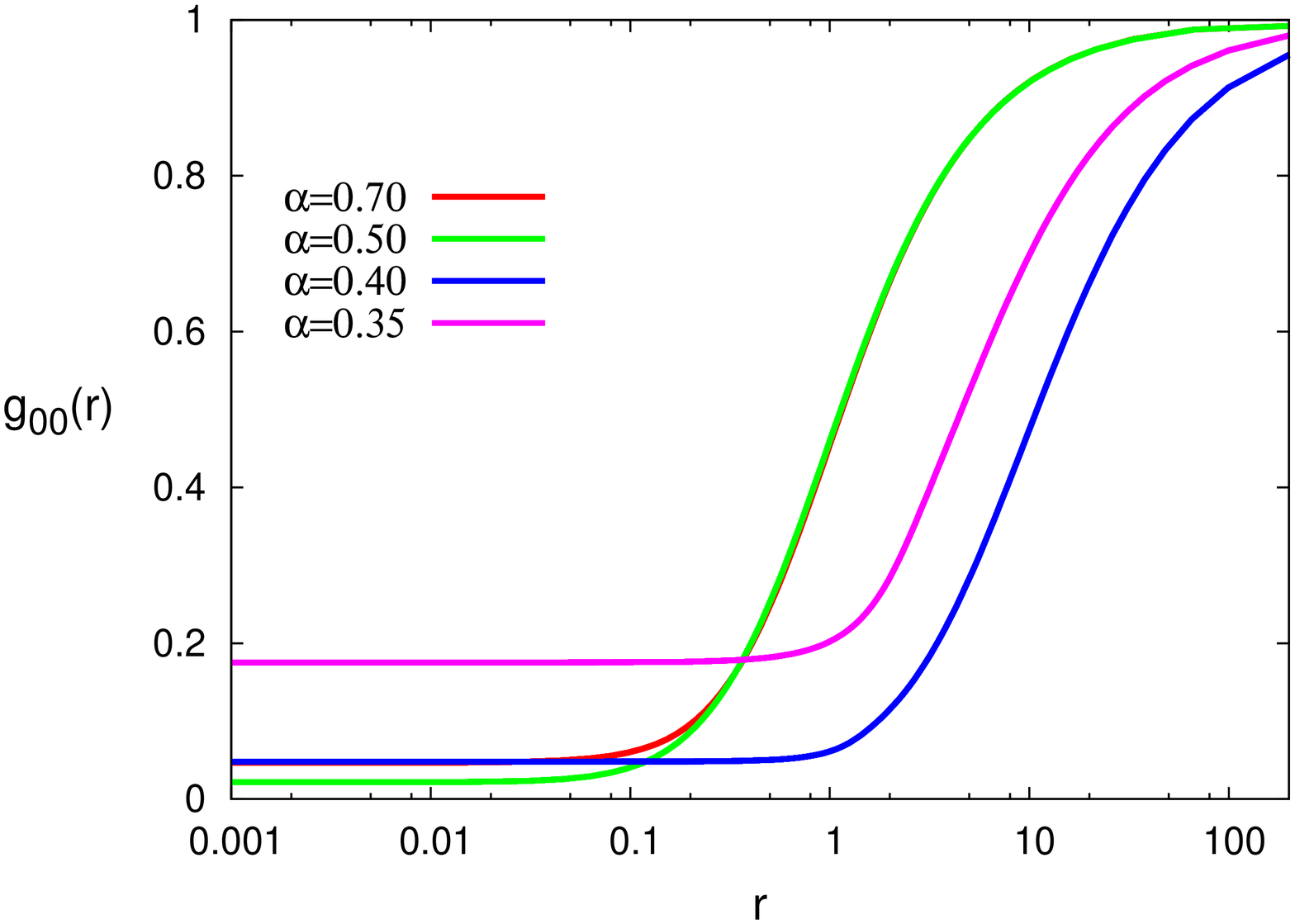}
\end{center}
\caption{\small Compact gauged Einstein-FLS boson stars on the second branch.
The profile functions of the scalar fields $X(r)$ and $Y(r)$ (upper plots), the gauge potential $A_0(r)$ (lower left), and the metric component $g_{00}(r)$  (lower right) are displayed as functions of the angular frequency $\omega$ for $g=0.15$, $\omega=0.75$ for a set of values of the gravitational coupling $\alpha$.}
    \lbfig{fig6}
\end{figure}

We expect that the evolution of the gauged boson stars in the model \re{lag} depends crucially on the ratio of the effective gravitational coupling $\alpha$ and the gauge coupling constant $g$. 
In particular, for small values of the effective gravitational coupling $\alpha$ and large values of the gauge coupling $g$, the two-branch structure observed in flat space should persist.  
Whereas for large values of the effective gravitational coupling $\alpha$ and small values of the gauge coupling $g$, the spiraling pattern should be present.
Thus, when gravity becomes strong enough, the correspondence with the usual scenario of the evolution of the boson stars should be recovered. 

Moreover, as the value of the mass parameter $\mu$ increases, the real component $\psi$ tends to its vacuum value everywhere in space, and the decoupled massive complex field $\phi$ satisfies the Einstein-Klein-Gordon equations. 
In this case the dependence of the mass and the charge of the gauged boson star on the angular frequency then possesses an infinite number of branches, representing the inspiraling of the solutions towards a limiting configuration %(see Fig.~\ref{fig2})
\cite{Friedberg:1986tp,Friedberg:1986tq,Kleihaus:2005me,Kleihaus:2007vk}.

Indeed, when we perform the calculations and scan the parameter space of the solutions we note that our observations agree with our expectations. 
%Note that this excludes the limiting case of a massless real component. 
In particular, in these calculations we vary either the gauge coupling $g$ or the gravitational coupling $\alpha$, while we keep all other parameters fixed. 
We start our presentation of the results for the case obtained by fixing the value of gravitational coupling $\alpha=0.3$ and varying the gauge coupling $g$, as demonstrated in Figs.~\ref{fig3} and \ref{fig4}.

Fig.~\ref{fig3} shows, that for relatively small gauge coupling $g$ the gauged boson stars show a spiraling behavior.
With increasing gauge coupling $g$, the mass and the charge of the gauged boson stars grow, including the respective values of the limiting (presumably singular) solution.
This pattern changes as the electrostatic repulsion becomes still stronger.
The gauged boson stars then show some intermediate behavior, reminiscent of the soliton boson stars with two maxima and a minimum in between, and the upper global maximum followed by a spiral.
However, now the second bifurcation point arises as the forward branch merges with the backward branch of radially excited boson stars \cite{Collodel:2017biu}.
At still larger values of the gauge coupling $g$ the branch pattern approaches the one of gauged Q-balls with basically two branches.

In all cases there is a minimal frequency $\omega_\mathrm{min}$, which increases with increasing gauge coupling $g$.
But the character of the second branch that bifurcates with the first one at $\omega_\mathrm{min}$ is very different, and strongly depends on the value of the gauge coupling $g$. 
In particular, mass and charge can either decrease or they can increase beyond the bifurcation point.
Moreover the second branch can either end at a second bifurcation point, or it can extend all the way to $\omega_\mathrm{max}$.

When we inspect the fields along the second branch we note that similarly to the corresponding solutions in Minkowski spacetime, the real field is vanishing in the interior region of the gauged boson star. 
Therefore the complex scalar field is massless there. 
Moreover, the electric potential is almost constant in the interior region.
The fields are displayed in Fig.~\ref{fig4} for configurations on the second branch for several values of the gauge coupling $g$, and constant values of the frequency $\omega$ and the gravitational coupling $\alpha$.
One clearly notes in the figure that configurations on the higher electrostatic branch (as illustrated in the figure for $g=0.1$) represent self-gravitating charged compactons.

For large values of the gauge coupling $g$, the characteristic size of the boson stars decreases slowly along the second branch.
The gravitational interaction makes them more compact. 
%If the gauge coupling is large enough, the second branch may extend up to the upper critical value of the angular frequency. 
For the intermediate range of values of $g$, %the second branch terminates at the second maximal angular frequency, as seen in the Fig.~\ref{fig3}. 
%Here it merges the new backward branch of gravitating solutions.
the size of compact boson stars continues to decrease also along the third branch. 
The minimum of the metric function $N(r)$ is then no longer located at the origin.
Instead it becomes associated with the position of the domain wall separating the interior of the configuration and its exterior. 

Of course, the corresponding pattern is observed when we fix the gauge coupling $g$ and scan the parameter space by varying the gravitational coupling $\alpha$.
This is illustrated in Figs.~\ref{fig5} and \ref{fig6}. 
Now for relatively small values of $\alpha$ the  two-branch scenario is recovered. 
However, when the gravitational coupling increases, first the second bifurcation point arises as the forward branch merges with the backward branch of radially excited boson stars, as seen in Fig.~\ref{fig5} for $g=0.15$ and $\alpha =0.35$.
(Furthermore, various multiboson star configuration may be linked to these excited branches \cite{Herdeiro:2021mol,Herdeiro:2020kvf}.) 
Finally, at large values of the gravitational coupling gravity takes over and the mini-boson star pattern is recovered.
%We remark that in the intermediate case, also the solutions on the third branch correspond to charged compactons.
We note that, as demonstrated in Fig.~\ref{fig5}, here the minimal frequency $\omega_\mathrm{min}$ first decreases with increasing gravitational coupling and then decreases again. 

%\textbf{Conjecture:} For each value of the gauge coupling $g$ there is a certain value of the gravitational coupling $\alpha$ for which the gravitational attraction stabilizes the electrostatic repulsion and, by analogy with the ungauged Q-ball in Minkowski spacetime, there will be just one branch of solutions extending down to the zero value of angular frequency.
%Alternatively, it may be linked to a RN black hole.
%\jk{I am not convinced of this conjecture.}

\subsection{Massless limit 
\boldmath {$\mu = 0$} \unboldmath}

%%%%%%%%%%%%%%%%%%%%%%%%%%%%%%%%%%%%%%%%%%%%%%%%%%%%%%%%%%%%%%%%%%%%%
\begin{figure}[t!]
\begin{center}
\includegraphics[height=.33\textheight,  angle =-90]{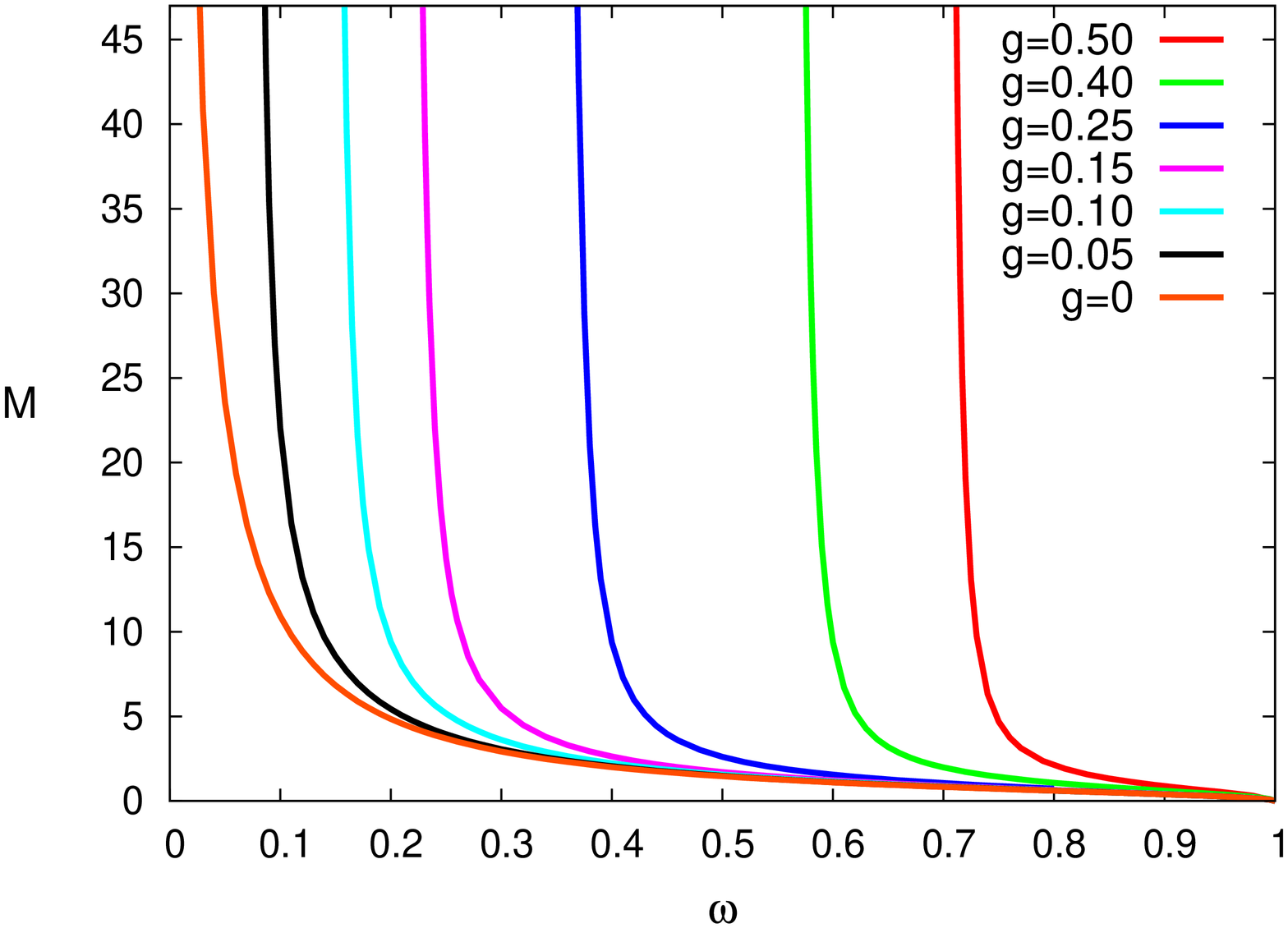}
\includegraphics[height=.33\textheight,  angle =-90]{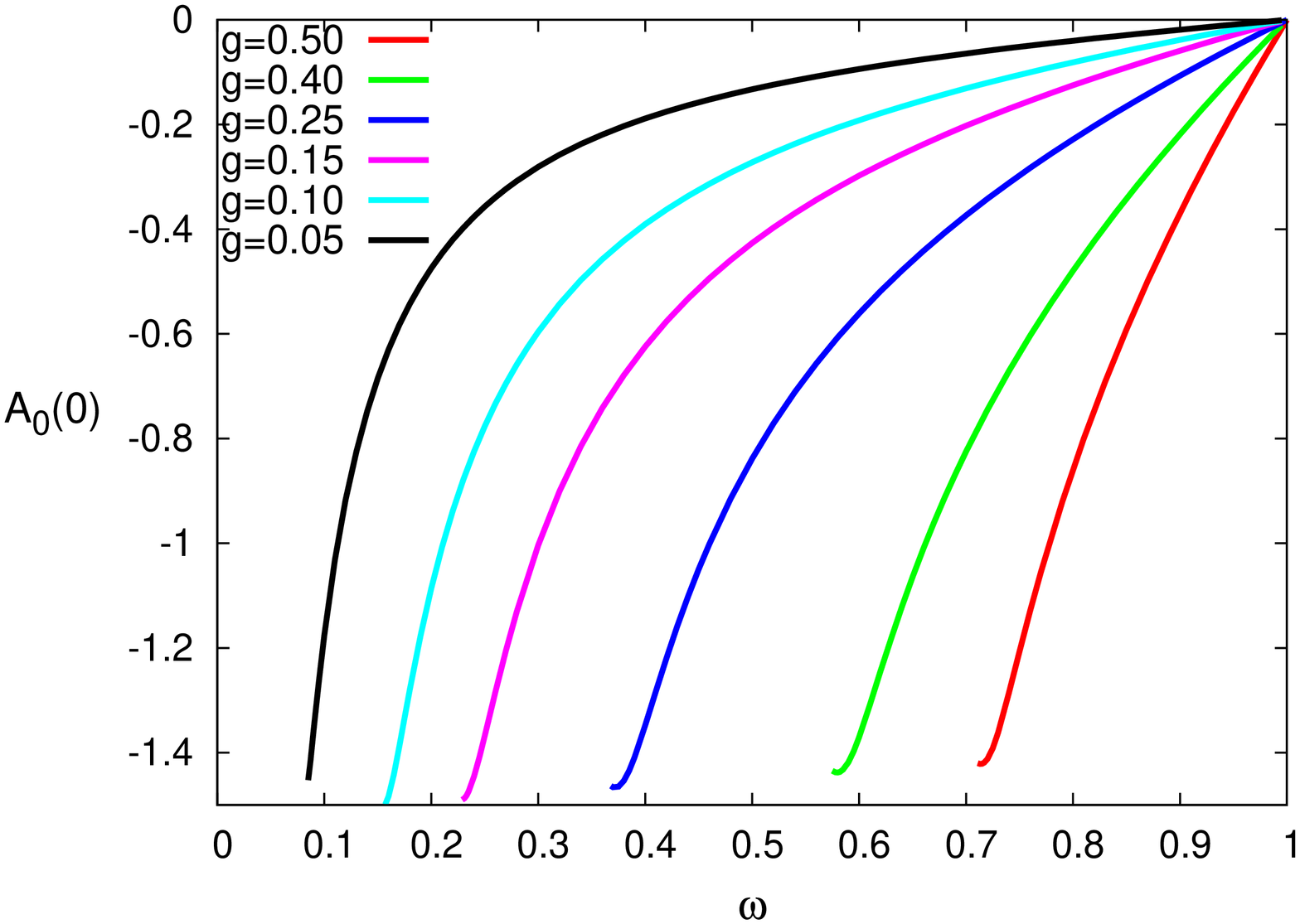}
\includegraphics[height=.33\textheight,  angle =-90]{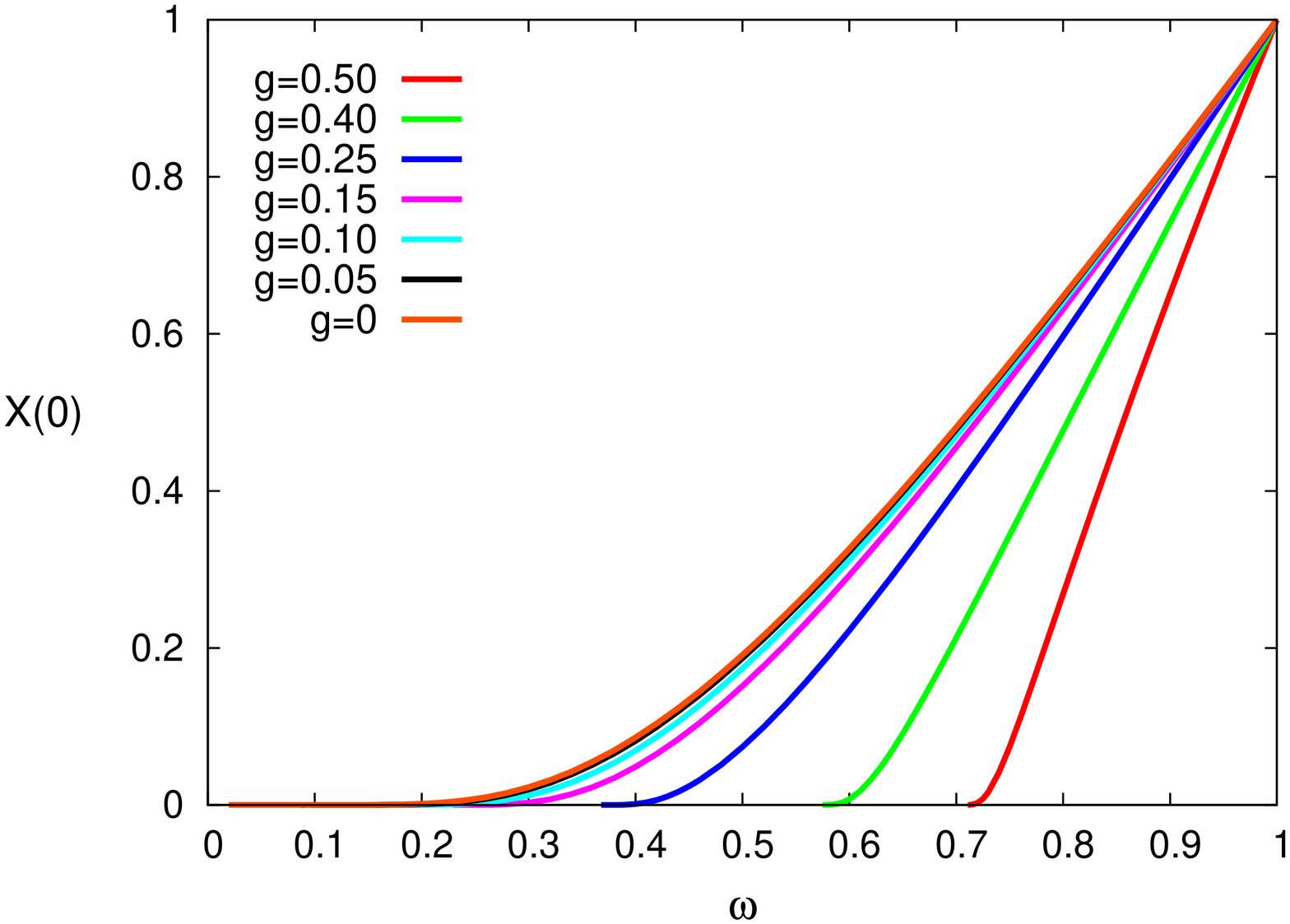}
\includegraphics[height=.33\textheight,  angle =-90]{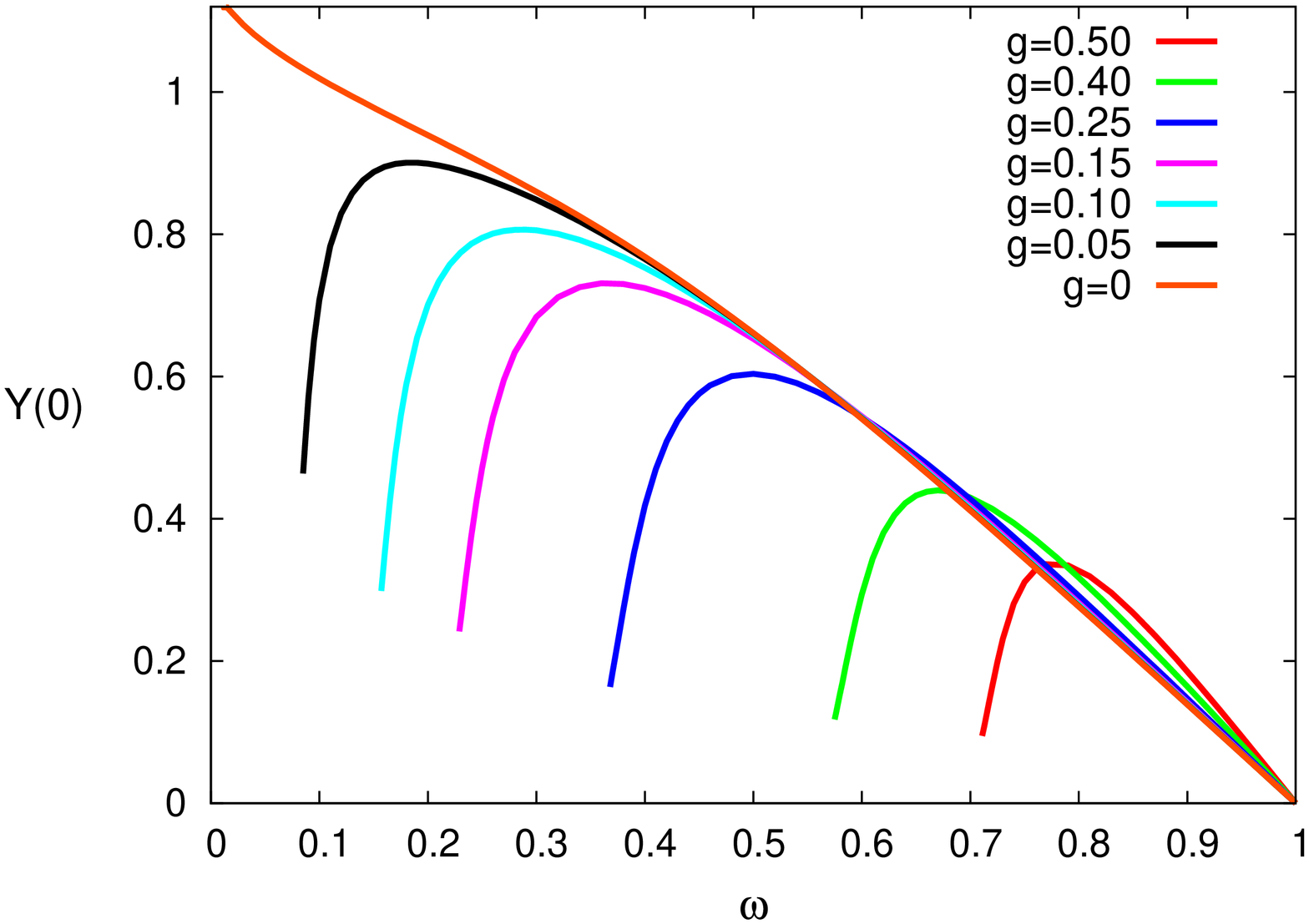}
\includegraphics[height=.33\textheight,  angle =-90]{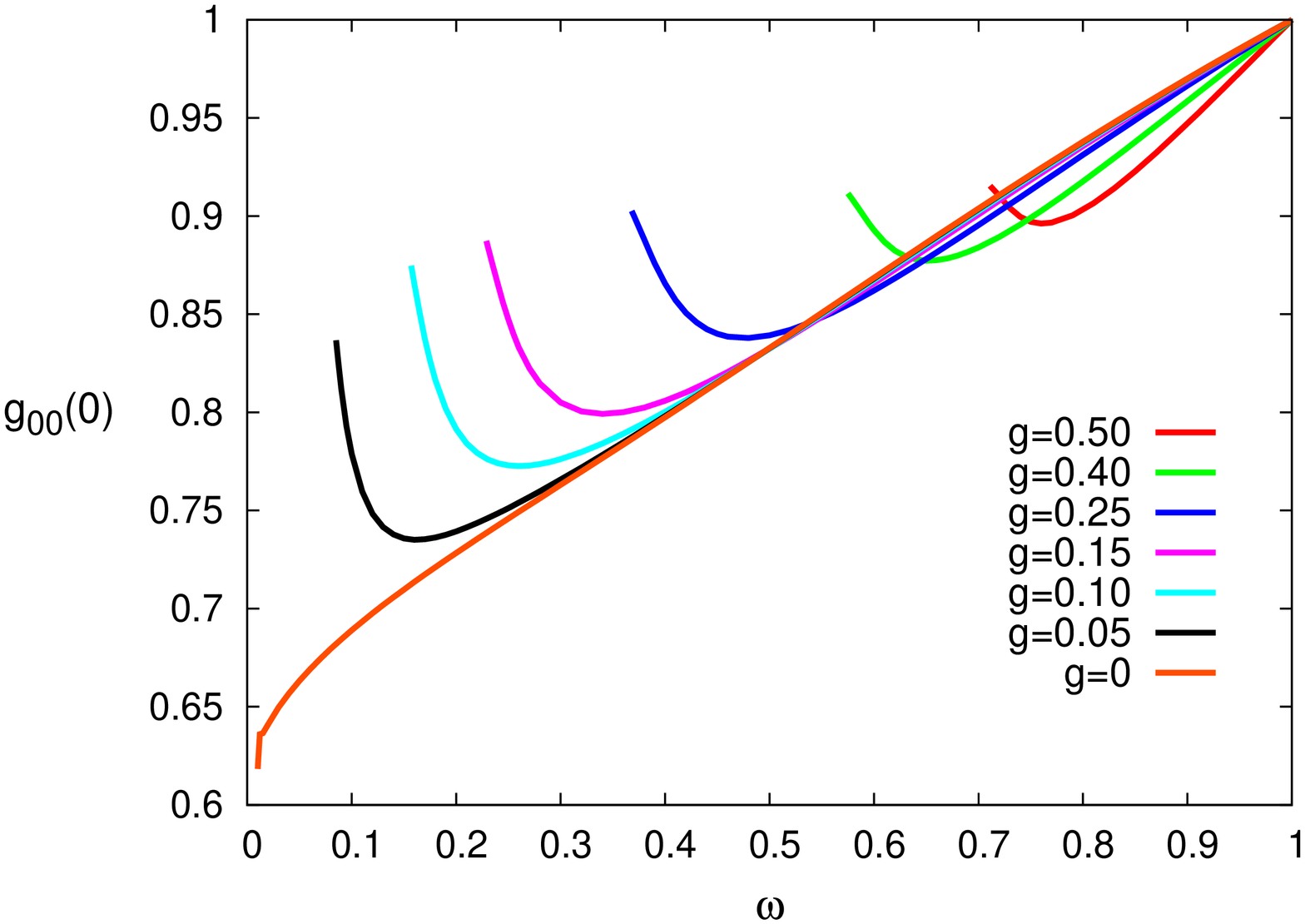}
\end{center}
\caption{\small Gauged Einstein-FLS boson stars in the massless limit $\mu=0$.
The energy of the solutions in units of $8 \pi$ (upper left plot), the values of the gauge potential $A_0$ (upper right plot), the scalar profile functions $X$, $Y$ (middle plots), and the metric component $g_{00}$ at $r=0$ (lower plot) are displayed as functions of the angular frequency $\omega$ for $\alpha=0.40$ for a set of values of the gauge coupling $g$.}
    \lbfig{fig7}
\end{figure}

%%%%%%%%%%%%%%%%%%%%%%%%%%%%%%%%%%%%%%%%%%%%%%%%%%%%%%%%%%%%%%%%%%%%%
\begin{figure}[t!]
\begin{center}
\includegraphics[height=.33\textheight,  angle =-90]{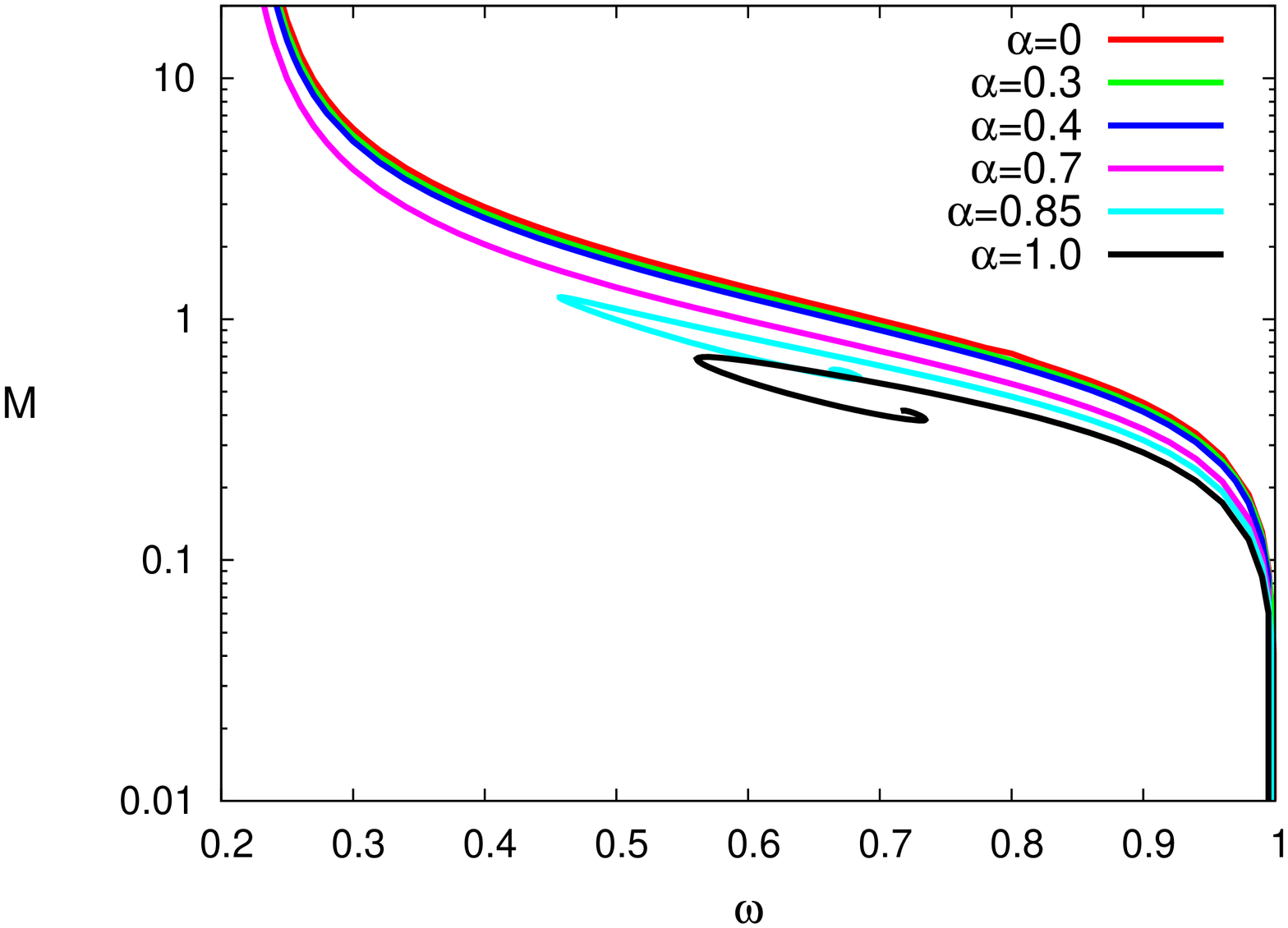}
\includegraphics[height=.33\textheight,  angle =-90]{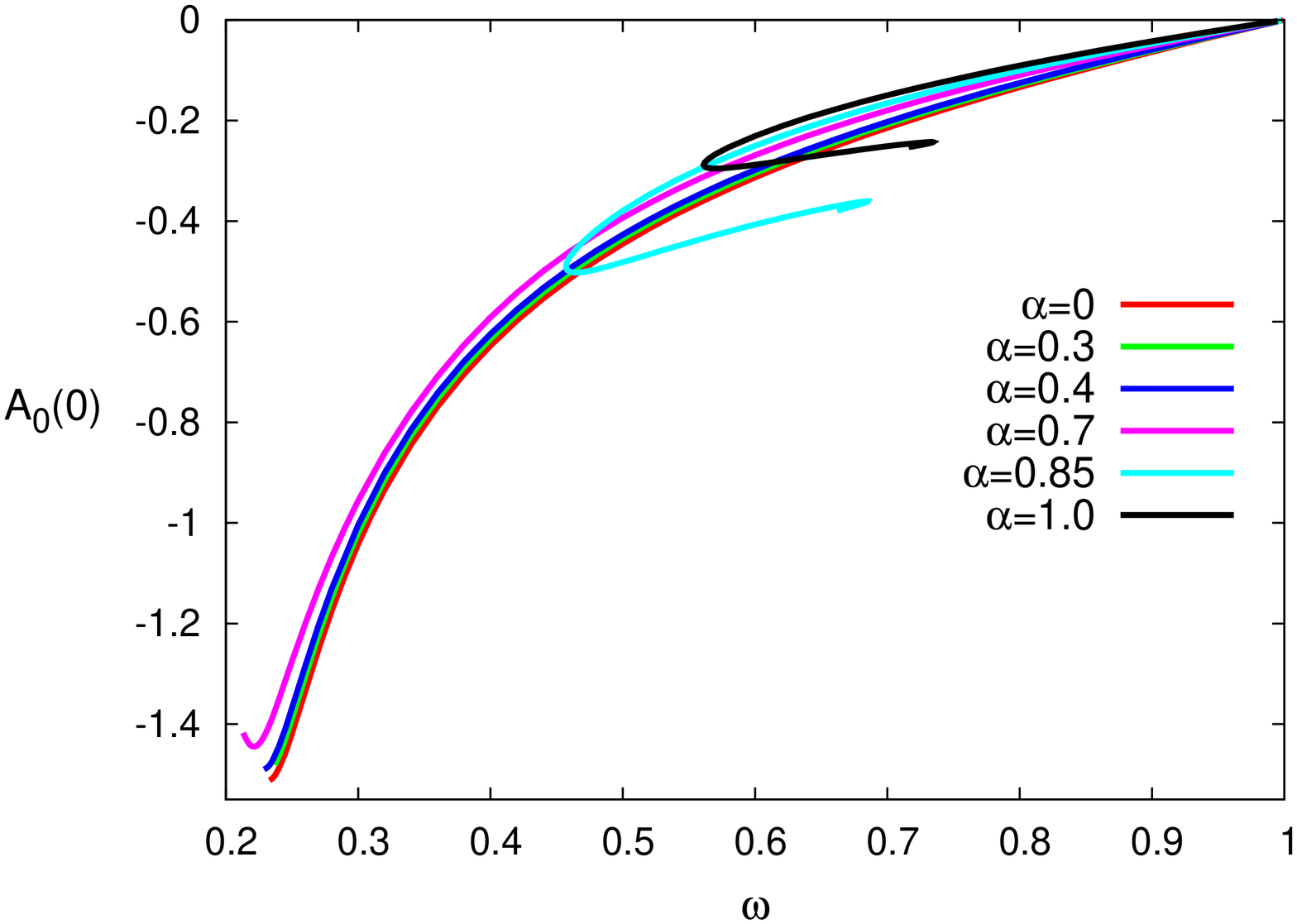}
\includegraphics[height=.33\textheight,  angle =-90]{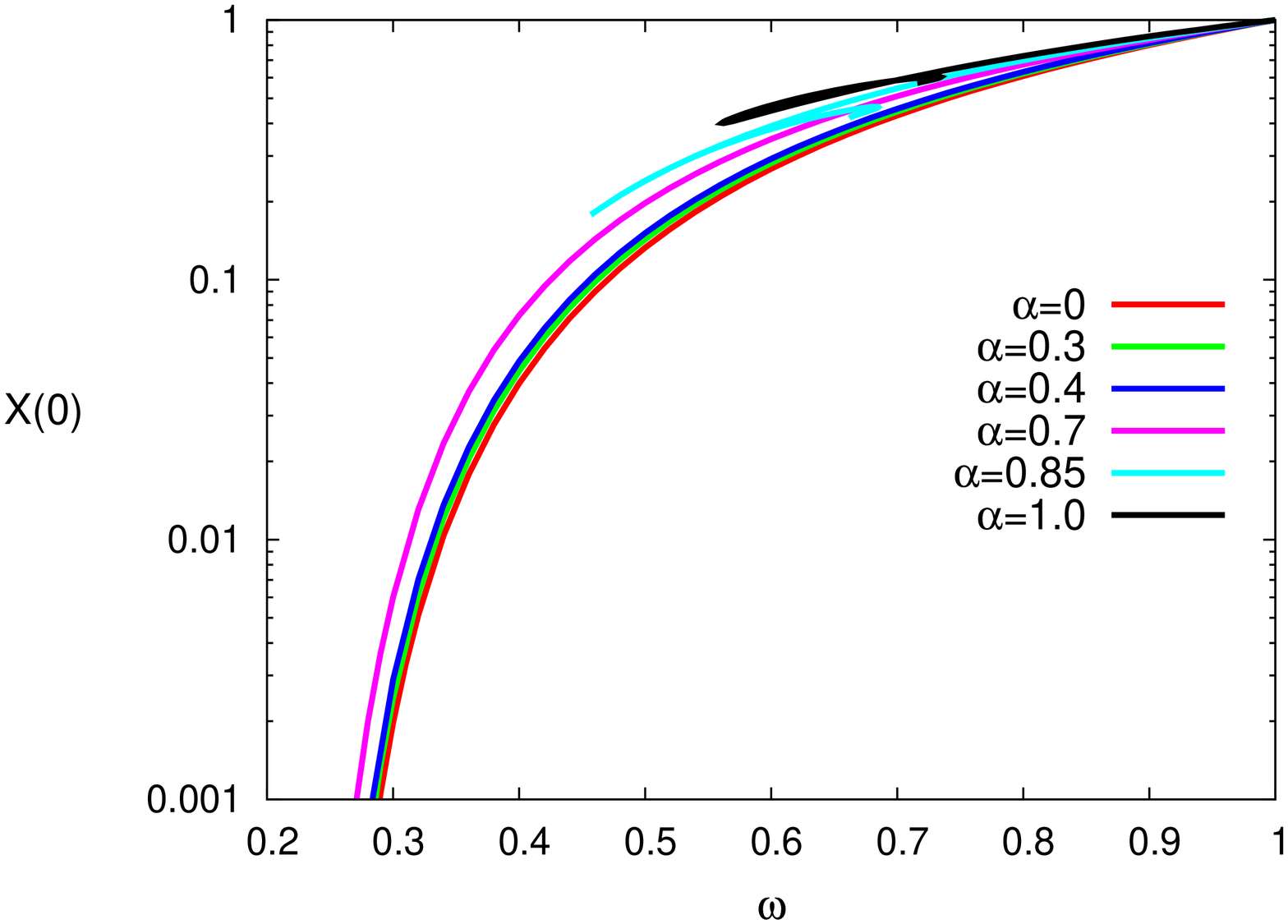}
\includegraphics[height=.33\textheight,  angle =-90]{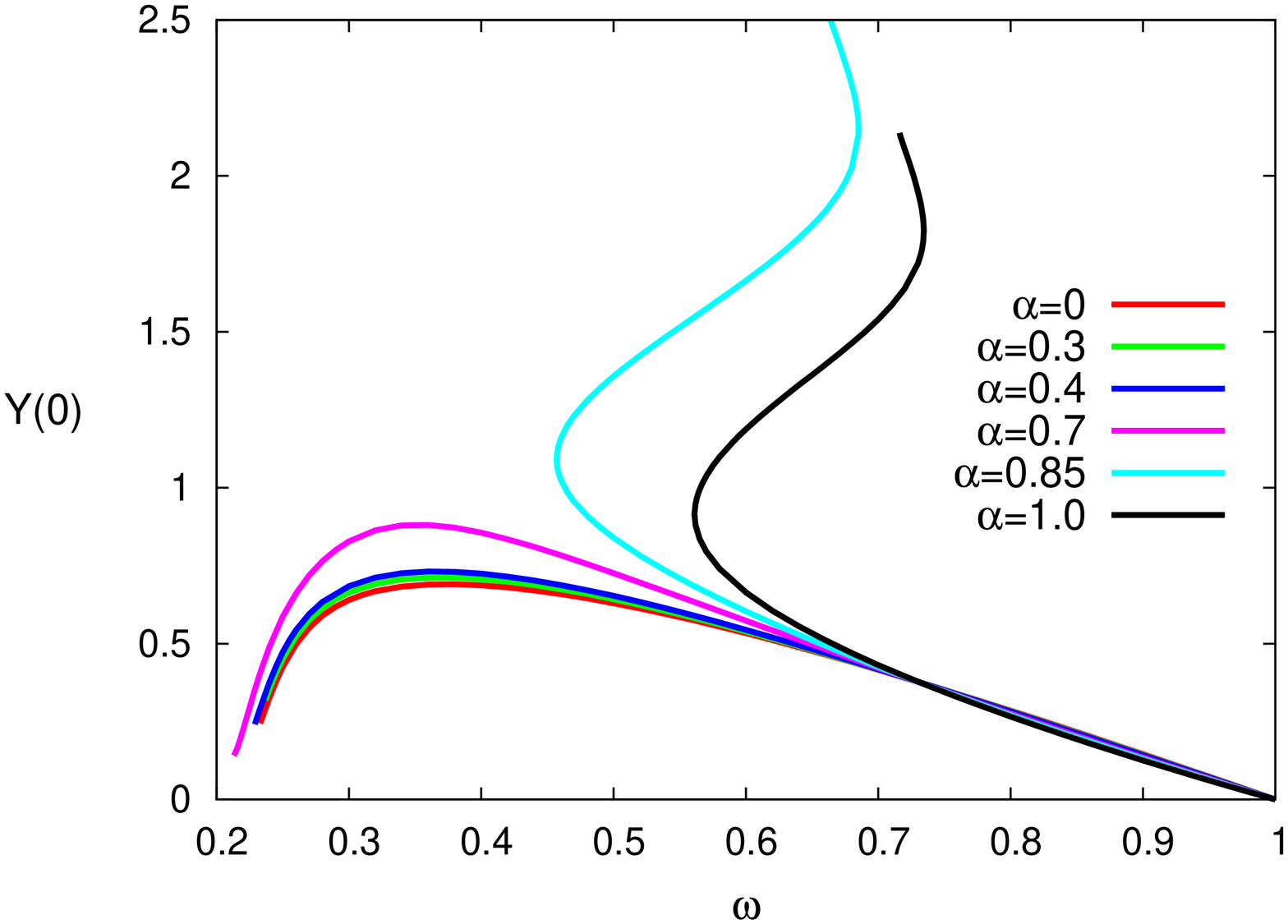}
\includegraphics[height=.33\textheight,  angle =-90]{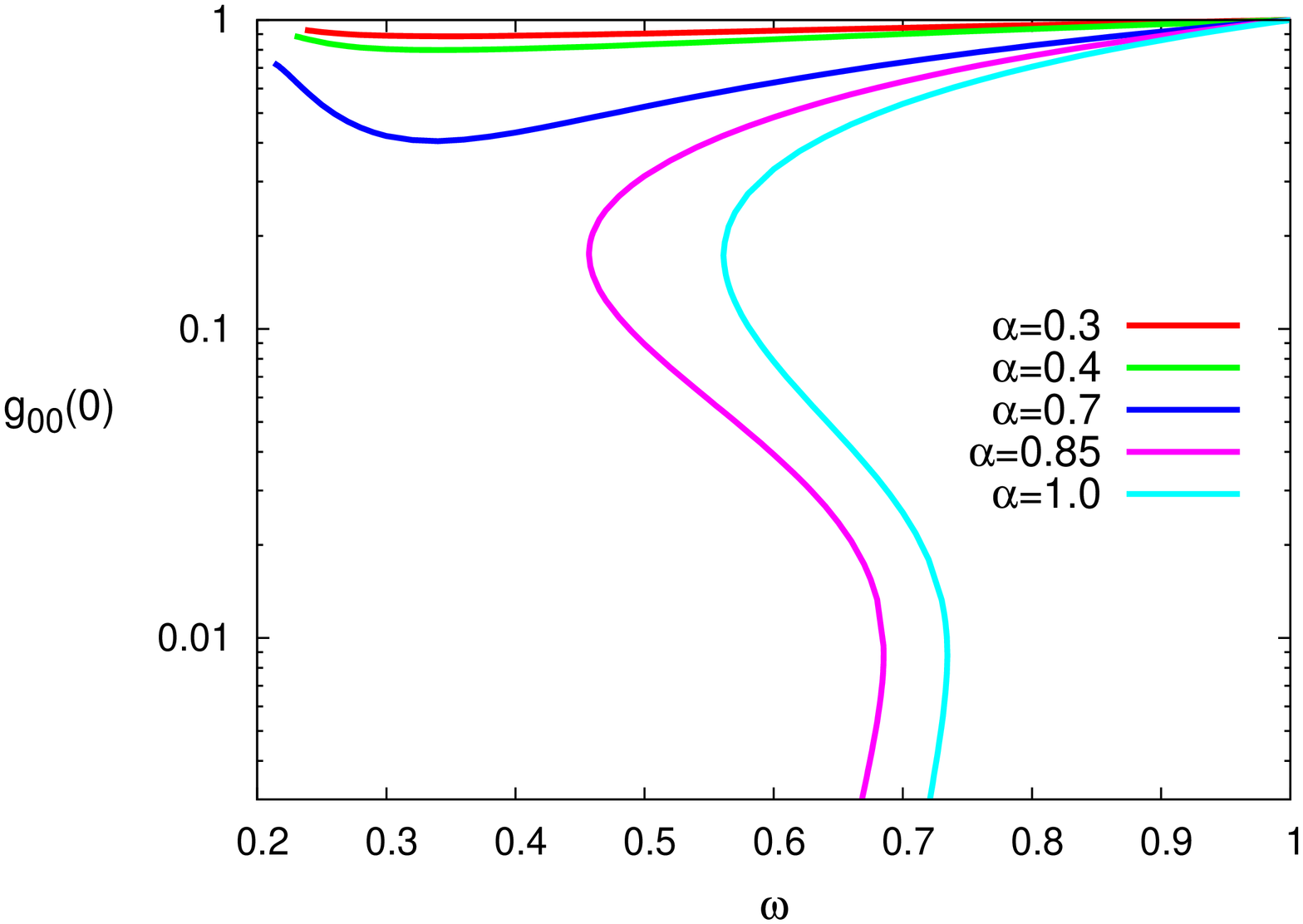}
\end{center}
\caption{\small Gauged Einstein-FLS boson stars in the massless limit $\mu=0$.
The energy of the solutions in units of $8 \pi$ (upper left plot), the values of the gauge potential $A_0$ (upper right plot), the scalar profile functions $X$, $Y$ (middle plots), and the metric component $g_{00}$ at $r=0$ (bottom plots) are displayed as functions of the angular frequency $\omega$ for $g=0.15$ for a set of values of the gravitational coupling $\alpha$.}
    \lbfig{fig8}
\end{figure}

We now turn to the $U(1)$ gauged boson stars in the limiting case $\mu=0$.
Like the Q-balls they possess a long range massless real scalar field \cite{Levin:2010gp,Loiko:2018mhb,Loiko:2019gwk}.
But the coupling to gravity yields another long range attractive interaction.
Since the gauged Q-balls exhibit very different properties in the case $\mu=0$, we expect to encounter corresponding differences also in the gravitating case, when constructing gauged boson stars in this limit.
Apart from that we expect again that the patterns found will be either dominated by the electromagnetic interaction or by gravity, depending on the relative strength of the respective coupling constants.

We start with a demonstration of the effect of the gauge coupling on boson stars by keeping the gravitational constant fixed at $\alpha=0.4$ and increasing the gauge coupling $g$.
We display in Fig.~\ref{fig7} the main characteristics of the self-gravitating charged boson stars obtained.
Note that in this case there still is only a single branch of solutions, which terminates at a minimal value of the angular frequency $\omega_\mathrm{min}$. 
Along the branch the mass and the charge of the configurations increase monotonically as $\omega$ decreases. 
Since the minimal frequency $\omega_\mathrm{min}$ increases with increasing gauge coupling $g$, the solutions will stop to exist beyond a maximal value of $g$.
Clearly, for $\alpha=0.4$ gravity is not yet sufficiently strong to dominate the properties of the gauged boson stars.
Consequently the pattern of the gauged Q-balls is retained.

We now inspect the behavior of the fields in Fig.~\ref{fig7} to gain a better understanding of the behavior of the solutions as the minimal frequency $\omega_\mathrm{min}$ is approached.
We note that toward this limit the massless real scalar field $\phi$ approaches zero within some region around the center of the configuration, such that the complex scalar field $\psi$ becomes also massless there. 
Consequently, a minor decrease of the angular frequency leads to a rapid inflation of the volume of the bubble. 
Then the balance between the volume energy and the surface energy becomes shifted and the gravitational interaction cannot stabilize the boson star any longer.

For larger values of the gravitational coupling $\alpha$, the usual spiral evolution of the boson stars is recovered, as long as the gauge coupling $g$ remains sufficiently small.
This is demonstrated in Fig.~\ref{fig8}, where we have chosen a small value for $g$ and increased the value of $\alpha$.
As expected, for small $\alpha$ the gauged boson stars are dominated by the electromagnetic interaction, and the pattern resembles the one discussed above.
However, as $\alpha$ becomes larger gravity takes over, and the usual boson star behavior with spiraling respectively oscillating behavior is indeed recovered.

\section{Conclusion}

We have considered Q-balls and boson stars in the Einstein-Maxwell-Friedberg-Lee-Sirlin model, where in addition to the gauged complex scalar field a real scalar field is present.
The real scalar field has a finite vacuum expectation value due to its quartic self-interaction potential. 
Consequently, the real scalar field carries a mass, that also depends on the strength of the self-interaction $\mu$.
For vanishing parameter $\mu$ the real scalar becomes massless and long-ranged. 
Independent of the value of $\mu$, the complex scalar field acquires its mass via its interaction with the real scalar field.

Gauged FLS Q-balls and boson stars possess quite distinct properties in the case of finite or vanishing parameter $\mu$, i.e., for short-ranged or long-ranged real scalar field
\cite{Friedberg:1976me,Levin:2010gp,Loiko:2018mhb,Loiko:2019gwk}.
In both cases an increase of the gauge coupling $g$ entails an increase of the minimal frequency $\omega_\mathrm{min}$ from zero, its value without the gauge field.
But for finite $\mu$ a bifurcation with a second branch of gauged Q-balls arises at $\omega_\mathrm{min}$, that extends to larger values of the frequency all the way to the maximal frequency $\omega_\mathrm{max}$,
whereas for vanishing $\mu$ there is no such second branch.

When gravity is coupled, another attractive interaction is present, and one has to consider the competition between gravity and electromagnetism. 
Depending on the relative strength of the respective coupling constants, the resulting gauged boson stars exhibit distinct behavior. 
When gravity dominates, the typical spiraling and oscillating pattern of boson stars is seen in their properties, when considered as functions of the frequency.
When electromagnetism dominates, however, the simple two branch structure ($\mu>0$) or single branch structure ($\mu=0$) of gauged FLS Q-balls is recovered.

Here we have only considered globally regular spherically symmetric configurations.
Interesting next steps will be on the one hand to address the expected associated hairy black holes and on the other hand to reduce the symmetry of the configurations and consider axially symmetric configurations as well as configurations with still less symmetry.
Spherically symmetric gauged FLS black holes are expected to exist in analogy to such black holes in the Einstein-Maxwell-scalar models
\cite{Herdeiro:2020xmb,Hong:2020miv,Brihaye:2021phs}.
Rotating axially symmetric boson stars and associated black holes have been constructed in the FLS model before \cite{Kunz:2019sgn}.
Here the inclusion of the Maxwell field will provide new interesting aspects. The work here should be taken further by considering boson stars in the extended $U(1)\times U(1)$ symmetric model \cite{Forgacs:2020vcy}. 
Moreover, the analogy to the states of the hydrogen atom observed in \cite{Herdeiro:2020kvf} would represent a fascinating endeavour for the FLS model as well.

\section*{Acknowledgment}
We are grateful to Carlos Herdeiro and Eugen Radu for inspiring and valuable discussions.
This work is supported by the DFG Research Training Group 1620 Models of Gravity, the COST Actions CA15117 and CA16104, and the Heisenberg-Landau program as well as the DAAD Ostpartnerschaftsprogramm. Ya.S. gratefully acknowledges the support by the Ministry of Education of Russian 
Federation, project No FEWF-2020-003.

 %%%%%%%%%%%%%%%%%%%%%%%%%%%%%%%%%%%%%%%%%%%%%%%%%%%%%%%%%%%%%%%%%%
 \begin{small}
 
%%%%%%%%%%%%%%%%%%%%%%%%%%%%%%%%%%%%%%%%%%%%%%%%%%%%%%%%%%%%%%%%%%%%%%%%%%%%%%
 \end{small}

 \end{document}